\documentclass[aps,prd,reprint,amsmath,amssymb,showpacs,preprintnumbers,superscriptaddress,nofootinbib]{revtex4-1}

\usepackage{slashed}
\usepackage{bm}
\usepackage{latexsym,amssymb,amsmath,float,url,mathrsfs}
\usepackage{bbold}
\usepackage{latexsym}
\usepackage{graphicx}
\usepackage{epstopdf}
\usepackage{amsmath}
\usepackage{subfigure}
\usepackage{feynmp}
\begin{document}



\title{Leptophilic Dark Matter with $Z'$ interactions}

\author{Nicole F.\ Bell} 
\affiliation{ARC Centre of Excellence for Particle Physics at the Terascale \\
School of Physics, The University of Melbourne, Victoria 3010, Australia}

\author{Yi Cai}
\affiliation{ARC Centre of Excellence for Particle Physics at the Terascale \\
School of Physics, The University of Melbourne, Victoria 3010, Australia}

\author{Rebecca K.\ Leane}
\affiliation{ARC Centre of Excellence for Particle Physics at the Terascale \\
School of Physics, The University of Melbourne, Victoria 3010, Australia}

\author{Anibal D.\ Medina}
\affiliation{ARC Centre of Excellence for Particle Physics at the Terascale \\
School of Physics, The University of Melbourne, Victoria 3010, Australia}
\date{\today}

\begin{abstract}
We consider a scenario where dark matter (DM) interacts exclusively
with Standard Model (SM) leptons at tree level.  Due to the absence of
tree-level couplings to quarks, the constraints on leptophilic dark
matter arising from direct detection and hadron collider experiments
are weaker than those for a generic WIMP.  We study a simple model in
which interactions of DM with SM leptons are mediated by a leptophilic
$Z'$ boson, and determine constraints on this scenario arising from
relic density, direct detection, and other experiments.  We then 
determine current LHC limits and project the future discovery reach.
We show that, despite the absence of direct interactions
with quarks, this scenario can be strongly constrained.
\end{abstract}

\maketitle

\section{Introduction}
\label{sec:intro}

\noindent

\noindent

The identity of dark matter (DM), thought to make up approximately 25
percent of the energy-matter content of the
universe~\cite{Kamionkowski_review, Bertone_review, Bergstrom_review},
has remained elusive since it was first suggested over 80 years ago.
Although there are an abundance of proposed particle physics models
which can potentially account for dark matter, the most well studied
class of model is that involving a weakly interacting massive particle
(WIMP).  WIMPs are appealing dark matter candidates because they can
naturally account for the dark matter relic abundance via the ``WIMP
miracle'' \cite{Bertone_review}, while also offering realistic
prospects for detection in a variety of direct detection, collider,
and indirect detection experiments.

However, the non-observation of WIMPs thus far has begun to place
meaningful constraints on the WIMP parameter space.  Experiments at
the LHC, which is a proton-proton collider, have not identified a DM
signal.  Direct detection (DD) experiments, which require nucleon-DM
interactions, similarly do not report a DM signal and are placing
tough limits on DM-nucleon cross-sections \cite{xenon:2013, lux:2014},
approaching the neutrino wall. However, the exclusions from these
experiments are based on DM-hadron interactions, perhaps hinting that
either the DM does not interact in such a way, or such interactions
are suppressed. To relax constraints on the DM parameter space, we
shall consider an alternate framework where direct DM-hadron
interactions do not occur, and instead the DM couples exclusively to
SM leptons at tree level. This is referred to as leptophilic dark
matter (LDM) \cite{cao:2009, ibarra:2009, Tait:2013, Fox:2009,
  Kopp:2009, Schmidt:2012, Agrawal:2014, Kopp:2014, Cohen:2009,
  Baltz:2002, Chen:2009, Spolyar:2009, Ko:2011, Chao:2011, Das:2013,
  Bi:2009, Bai:2014}.

\begin{figure}[h]
\begin{fmffile}{loop2}
\begin{fmfgraph*}(150,55)
\fmfset{arrow_len}{8}
\fmfkeep{loop}
\fmfleft{i1,i2}
\fmfright{o1,o2}
\fmflabel{$\chi$}{i1}
\fmflabel{$\bar{\chi}$}{i2}
\fmflabel{$N$}{o1}
\fmflabel{$N$}{o2}
\fmf{fermion,tension=.5}{i1,v1,i2}
\fmf{fermion,tension=.5}{o1,v3,o2}
\fmfblob{.1w}{v1}
\fmf{photon,label=$\gamma$}{v2,v3}
\fmf{fermion,right,tension=.35,label=$\ell$}{v1,v2,v1}
\end{fmfgraph*}
\end{fmffile}
\vspace{4 mm}
\caption{Loop-suppressed direct detection signal for leptophilic dark
  matter.}
\label{fig:loop2}
\end{figure}
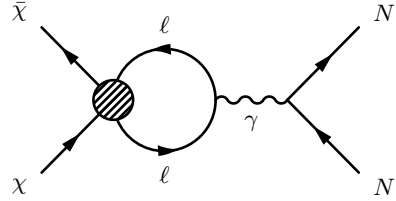

A leptophilic DM model may be tested via the three usual DM searches:
direct detection, indirect detection and collider searches.  However,
the phenomenology of leptophilic DM is quite different to that of a
generic WIMP.  Although leptophilic DM does not couple to quarks at
tree-level, such couplings will inevitably exist at higher order.  DM
detection processes involving quarks will thus still yield limits,
though they will be relaxed by the presence of loop factors and/or
additional coupling constants, thereby increasing the allowed
parameter space.  For example, there exists a loop-suppressed direct
detection process shown in Fig.~\ref{fig:loop2} \cite{Kopp:2009}.  We shall
see that despite the suppressed nature of this process, direct
detection still places meaningful constraints on leptophilic dark
matter.

\hspace{12 mm}

Indirect detection experiments are a scan of the astrophysical sky,
searching for unexplained fluxes which may be a produced as a result
of DM annihilation or decay.  For low DM masses, interesting indirect
detection bounds have been placed using Fermi observations of dwarf
spheroidal galaxies \cite{fermi:2014}.  At higher DM masses,
significant interest in leptophilic DM was sparked by an excess in the
cosmic ray positron fraction measured by the PAMELA
\cite{pamela:2013}, Fermi \cite{fermi:2012} and AMS~\cite{AMS:2013}
experiments.  Given that no corresponding antiproton excess was seen,
this suggested models in which DM annihilates, with a large cross
section, to leptonic final states.  However, these signals are subject
to significant astrophysical uncertainties, and may in fact be
produced by nearby pulsars \cite{Hooper:2008, Profumo:2008}.
Moreover, as noted above, a model can only be ``leptophilic'' at
lowest order with higher order process inevitably producing hadrons
with non-negligible fluxes \cite{Kachelreiss:2009}.  In this work, we
shall not attempt to explain the positron results, but instead focus
on LHC and direct detection bounds on leptophilic dark matter.

Collider searches at the LHC and LEP \cite{Fox:2011, Kopp:2014} have
placed interesting limits on standard WIMPs.  The most generic,
model independent limits are those obtained from the mono-X searches
\cite{CMS:2012, ATLAS1:2013, monolep:2013, monolep:2013, monoz:2012,
  bell:2012, ATLAS2:2014, ATLAS2:2013} (mono-jet, mono-photon or
mono-W/Z).  These signals are obtained when a single SM particle
recoils against missing momentum, attributed to dark matter particles
which escape undetected.  However, the mono-X searches at the LHC
require that dark matter to couples to quarks, and are thus not
applicable for a leptophilic scenario.  There exist mono-photon limits
from LEP \cite{Fox:2011}, however these are only relevant for DM
coupling to electrons (rather than muons or taus) and only extend to
low DM masses.  The LHC collider signals for leptophilic DM are very
different from the mono-X signals and will be discussed in section
\ref{sec:lhc} below in the context of our example model.

The outline of the paper is as follows: We introduce a simple
leptophilic model in Section~\ref{sec:model}, consider relic density
requirements in Section~\ref{sec:relic}, and discuss direct detection
and other constraints in Section~\ref{sec:direct} and
\ref{sec:constraints} respectively.  The LHC collider phenomenology
for leptophilic dark matter is discussed in Section~\ref{sec:lhc},
and our main results are summarized in
Fig.~\ref{Fig:VV}, \ref{fig:mueall22}, \ref{fig:mueall2} and
\ref{fig:mueall}.

\section{Leptophilic Model}
\label{sec:model}

To fully investigate a LDM scenario, we shall adopt a particular simple model.
We introduce a new spin-1 vector boson, $Z'$, which mediates interactions between SM leptons and the DM. Such a setup can be described via the Lagrangian
\begin{eqnarray} \label{eq:lagr1}
 \mathcal{L} &=& \mathcal{L}_{SM}-\frac{1}{4}{Z}'_{\mu\nu}{Z}'^{\mu\nu} - \frac{\epsilon}{2}{Z}'_{\mu\nu}{B}^{\mu\nu} + i{\bar{\chi}}\gamma_\mu\partial^\mu\chi \\ \nonumber
&&+ {\bar{\chi}}\gamma^\mu(g_\chi^V+g_\chi^A\gamma^5)\chi {Z}'_\mu 
+ {\bar{\ell}}\gamma^\mu(g_\ell^V+g_\ell^A\gamma^5)\ell {Z}'_\mu \\ && - m_\chi \bar{\chi}\chi + \frac{1}{2}m_{Z'}^2 Z'_\mu Z'^\mu, \nonumber \end{eqnarray}
where $\epsilon$ is the kinetic mixing parameter of $Z'$ and SM
hypercharge gauge boson, $\ell = e, \mu, \tau, \nu_e, \nu_\mu,
\nu_\tau$ are the SM leptons, $g_\ell=g_e,g_\mu,g_\tau$ are the $Z'$
coupling strengths to each SM lepton flavor, and $g_\chi$ is the
coupling strength of the $Z'$ to DM.  We allow both vector (V) and
axial-vector (A) couplings of the $Z'$.  The parameters to investigate,
therefore, are $g_\chi, g_\ell, m_{Z'}, m_{\chi}$, along with relevant
cross sections. In this general setup, a mass generation mechanism for
the $Z'$ and DM is not specified. 

At low energies, such as those relevant for direct detection, the $Z'$
interactions can be well approximated by an effective contact
operator,
\begin{equation}
\mathcal{L}_{eff}=\frac{1}{\Lambda^2}(\bar{\chi}\Gamma_\chi\chi)(\bar{\ell}\Gamma_\ell\ell),
\label{eq:eff}
\end{equation}
where the effective cut-off scale is
\begin{equation}
\Lambda=\frac{m_{Z'}}{\sqrt{g_\chi g_\ell}}.
\label{eq:lambda}
\end{equation}
Given the form of the $Z'$ interactions in Eq. \ref{eq:lagr1}, the
possible Lorentz structures are combinations of vector (V) and
axial-vector (A) bilinears: $\Gamma_\chi\otimes\Gamma_\ell$ = $V\otimes
V$, $A\otimes V$, $V\otimes A$ or $A\otimes A$.  However, in order to
permit SM Yukawa couplings without breaking $U(1)_L$ gauge invariance,
the $Z'$ coupling to the SM leptons must be vectorlike, thus we shall
require $\Gamma_\ell$ = $V$.  We list the possible Lorentz structures
in Table~\ref{tab:lorentz}, and summarise their pertinent features.

\renewcommand{\arraystretch}{1.2} 
\begin{table}[htb]
  \begin{center}
   \caption{Lorentz structure of the $Z'$ couplings.  For axial vector
     couplings to leptons, the loop-level direct detection signal
     vanishes~\cite{Kopp:2009}.  However, in order for the SM Yukawa couplings to
     respect $U(1)_L$ gauge invariance, the $Z'$ couplings to leptons
     should be vectorlike.  Also note that $\Gamma_\chi = V$ is not
     permitted for Majorana DM.\label{tab:lorentz}}
       \begin{tabular}{|c|c|c|c|}
      \hline
        $\Gamma_\chi\otimes\Gamma_\ell$ &  $\sigma(\chi\chi\rightarrow \overline{\ell} \ell)$ & $\sigma(\chi N \rightarrow \chi N)$ & Gauge invariant? \\
      \hline\hline
        $V\otimes V$     &   $s$-wave &  1 (1-loop)     & Yes  \\
        $A\otimes V$     &   $p$-wave &  $v^2$ (1-loop) &  Yes \\
        $V\otimes A$     &   $s$-wave &  -              &  No  \\
        $A\otimes A$     &   $p$-wave &  -              &  No  \\
    \hline
    \end{tabular}
\end{center}
\end{table}

There has been some previous work on vector bosons which couple only
to leptons.  Considering only SM fields, the symmetries $U(1)_{L_i -
  L_j}$ are anomaly free and can thus be gauged \cite{Foot:1990,
  Volkas:1990, He:1991}; models in which DM interacts via a
$L_\mu-L_\tau$ gauge boson have recently been explored in
\cite{Das:2013, Baek:2009, Bi:2009}.  We take a different approach in
the present work and consider a leptophilic $Z'$ which couples to a
single lepton flavor, taking each flavour in turn.
Phenomenologically, this is a natural choice because the experimental
constraints depend on which lepton flavor is being considered.  With
this approach, new dark sector particles must be chosen with the
correct quantum numbers to cancel anomalies.  However, any hidden
sector particles other than the DM candidate can be taken to be heavy
enough to be decoupled from our Lagrangian, while light enough to
still contribute to anomaly cancellation. We thus take the coupling
strengths $g_\ell^{V}$ and $g_\chi^{A,V}$ as free parameters to be
constrained.

\section{Dark Matter Relic Density}
\label{sec:relic}

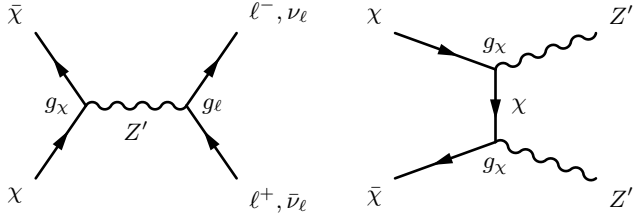
\begin{figure}[h]
\centering
\begin{subfigure}
  \centering
\begin{fmffile}{diagrama}
\begin{fmfgraph*}(95,55)
\fmfset{arrow_len}{8}
\fmfleft{i1,i2}
\fmfright{o1,o2}
\fmflabel{$\chi$}{i1}
\fmflabel{$\bar{\chi}$}{i2}
\fmflabel{$\ell^+,\bar{\nu}_\ell$}{o1}
\fmflabel{$\ell^-,\nu_\ell$}{o2}
\fmflabel{$g_\chi$}{v1}
\fmflabel{$g_\ell$}{v2}
\fmf{fermion}{i1,v1,i2}
\fmf{fermion}{o1,v2,o2}
\fmf{photon,label=$Z'$}{v1,v2}
\end{fmfgraph*}
\end{fmffile}
\end{subfigure}\hfill
\hspace{8 mm}
\begin{subfigure}
  \centering
\begin{fmffile}{diagramz}
\begin{fmfgraph*}(95,55)
\fmfset{arrow_len}{8}
\fmfleft{i1,i2} 
\fmfright{o1,o2}
\fmflabel{$\bar{\chi}$}{i1}
\fmflabel{$\chi$}{i2}
\fmflabel{$g_\chi$}{v1}
\fmflabel{$g_\chi$}{v2}
\fmf{fermion}{v1,i1} 
\fmf{boson}{v1,o1}
\fmf{fermion}{i2,v2}
\fmf{boson}{v2,o2}
\fmflabel{$Z'$}{o1}
\fmflabel{$Z'$}{o2}
\fmf{fermion,label=$\chi$}{v2,v1} 
\end{fmfgraph*}
\end{fmffile}
\end{subfigure}\hfill
\vspace{4 mm}
\caption{DM annihilation processes, which
  determine the relic density at freezeout.}
\label{fig:relden}
\end{figure}

In our leptophilic framework the dominant DM annihilation
channels are
 \begin{equation}
\chi\bar{\chi} \rightarrow \ell^-\ell^+, \bar{\nu}_\ell\nu_{\ell}, Z'Z',
 \end{equation}
with the corresponding Feynman diagrams shown in
Fig.~\ref{fig:relden}.  To determine the parameter space allowed by
relic density constraints, we implemented our model with FeynRules
\cite{Feyn:2009} and generated model files for MicroMEGAs
\cite{Micro:2013} for the relic density calculation.  We then
performed a scan over $m_{Z'}, m_\chi, g_\ell, g_\chi$ to determine the
parameters which yield the correct relic density, $\Omega_\chi h^2 =
0.1187\pm0.0017$~\cite{Planck:2013}.

\vspace{4 mm}

In Fig. \ref{Fig:VV}, \ref{fig:mueall22}, \ref{fig:mueall}, we plot relic density curves in the $g_\ell=g_\chi$
vs $m_{Z'}$ plane, for various choices of the DM mass, and in Fig. \ref{fig:mueall2} for $4g_\mu=g_\chi$ and $8g_\mu=g_\chi$.  Parameters
must lie on these curves to produce the correct relic density.  Larger
values of the couplings would lead to a subdominant contribution to
the relic density, while smaller couplings would overclose the
universe unless additional annihilation channels were present.

The features of the relic density curves can be easily understood: The
$Z'Z'$ channel is kinematically open only for $m_{Z'} < m_\chi$, while
for $m_{Z'} > m_\chi$ the freeze-out is determined by annihilation to
leptons.  The annihilation cross section to leptons has an $s$-wave
contribution when $\Gamma_\chi=V$, but proceeds via a velocity
suppressed $p$-wave contribution when $\Gamma_\chi=A$.  Resonant
production of DM occurs when $m_{Z'}\approx2m_\chi$, seen as strong
dips in the relic density curves.

\section{Direct Detection}
\label{sec:direct}

\begin{figure}[h]
\begin{fmffile}{loop}
\begin{fmfgraph*}(150,55)
\fmfset{arrow_len}{8}
\fmfkeep{loop}
\fmfleft{i1,i2}
\fmfright{o1,o2}
\fmflabel{$\chi$}{i1}
\fmflabel{$\bar{\chi}$}{i2}
\fmflabel{$N$}{o1}
\fmflabel{$N$}{o2}
\fmflabel{$g_\chi$}{v1}
\fmf{fermion,tension=.5}{i1,v1,i2}
\fmf{fermion,tension=.5}{o1,v4,o2}
\fmf{photon,label=$Z'$}{v1,v2}
\fmf{photon,label=$\gamma$}{v3,v4}
\fmf{fermion,right,tension=.35,label=$\ell$}{v2,v3,v2}
\end{fmfgraph*}
\end{fmffile}
\vspace{4 mm}
\caption{Dominant direct detection process with a leptophilic $Z'$ as SM-DM mediator.}
\label{fig:loopa}
\end{figure}
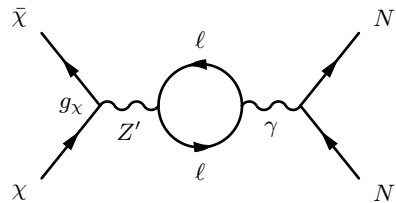

Direct detection experiments measure the recoil energy of SM nuclei
after DM scattering. There are two ways leptophilic DM can be found at
DD experiments: scattering at tree-level with electrons (particularly
for experiments which do not veto leptonic recoils (i.e. DAMA/LIBRA)),
or at higher orders with quarks. For tree-level scattering, the DM
will scatter with electrons bound to atoms, causing electrons to
absorb the recoil energy and eject from the atom, or remain bound to
the atom but with increased energies and recoiled nuclei. It turns out
this electron scattering is not very energetic, so that it cannot
usually be seen at DD experiments. In the rare case the electron has a
high enough initial momentum to produce a sufficiently energetic DD
signal, the cross section suffers from large wavefunction
suppression. Because of this, loop suppressed scattering with quarks
will typically dominate DD for leptophilic DM \cite{Kopp:2009}
provided the diagrams do not vanish. The dominant loop level process for
direct detection in our $Z'$ model is shown in Fig. \ref{fig:loopa}.
A similar diagram in which the photon is replaced by a $Z$ makes a
negligible contribution.

\hspace{12 mm}

The loop suppressed contributions to the DD cross section were
calculated in the effective operator framework in
Ref.~\cite{Kopp:2009}.  For the Lorentz structures
$\Gamma_\chi\otimes\Gamma_\ell=A\otimes A$ and $V\otimes A$, the
diagram in Fig.~\ref{fig:loop2} vanishes, and thus direct detection
provides no constraint.  However, these Lorentz structures are not
interesting for our model, because we require $\Gamma_\ell=V$ in order
for the SM Yukawa couplings to respect $U(1)_L$ gauge invariance.  For
the remaining possibilities, $\Gamma_\chi\otimes\Gamma_\ell=V\otimes
V$ and $A\otimes V$, the DM-nucleus cross sections are given by
\begin{eqnarray}
&&\sigma_{VV}=\frac{\mu_N^2}{9\pi}\left[\frac{\alpha_{EM}Z}{\pi\Lambda^2}\textrm{log}\left(\frac{m_\ell^2}{\mu^2}\right)\right]^2,
\label{eq:d1}
\\
&&\sigma_{AV}=\frac{\mu_N^2v_\chi^2}{9\pi}\left(1+\frac{\mu_N^2}{2m_N^2}\right)\left[\frac{\alpha_{EM}Z}{\pi\Lambda^2}\textrm{log}\left(\frac{m_\ell^2}{\mu^2}\right)\right]^2,
\label{eq:d2}
\end{eqnarray}
where $v_\chi^2\approx10^{-6}$ is the DM velocity, $m_N$ and $Z$ are
the target nucleus' mass and charge respectively,
$\mu_N=m_Nm_\chi/(m_N+m_\chi)$ is the reduced mass of the DM-nucleus
system and $\mu$ is the renormalization scale.  As in \cite{Kopp:2009}
we set $\mu=\Lambda$.\footnote{As the cross sections depend only
  logarithmically on $\mu$, extreme choices for this parameter can
  change the limits on $g_{\ell,\chi}$ by at most a factor of a few.}

\begin{figure}[h]
  \centering
  \includegraphics[width=\columnwidth]{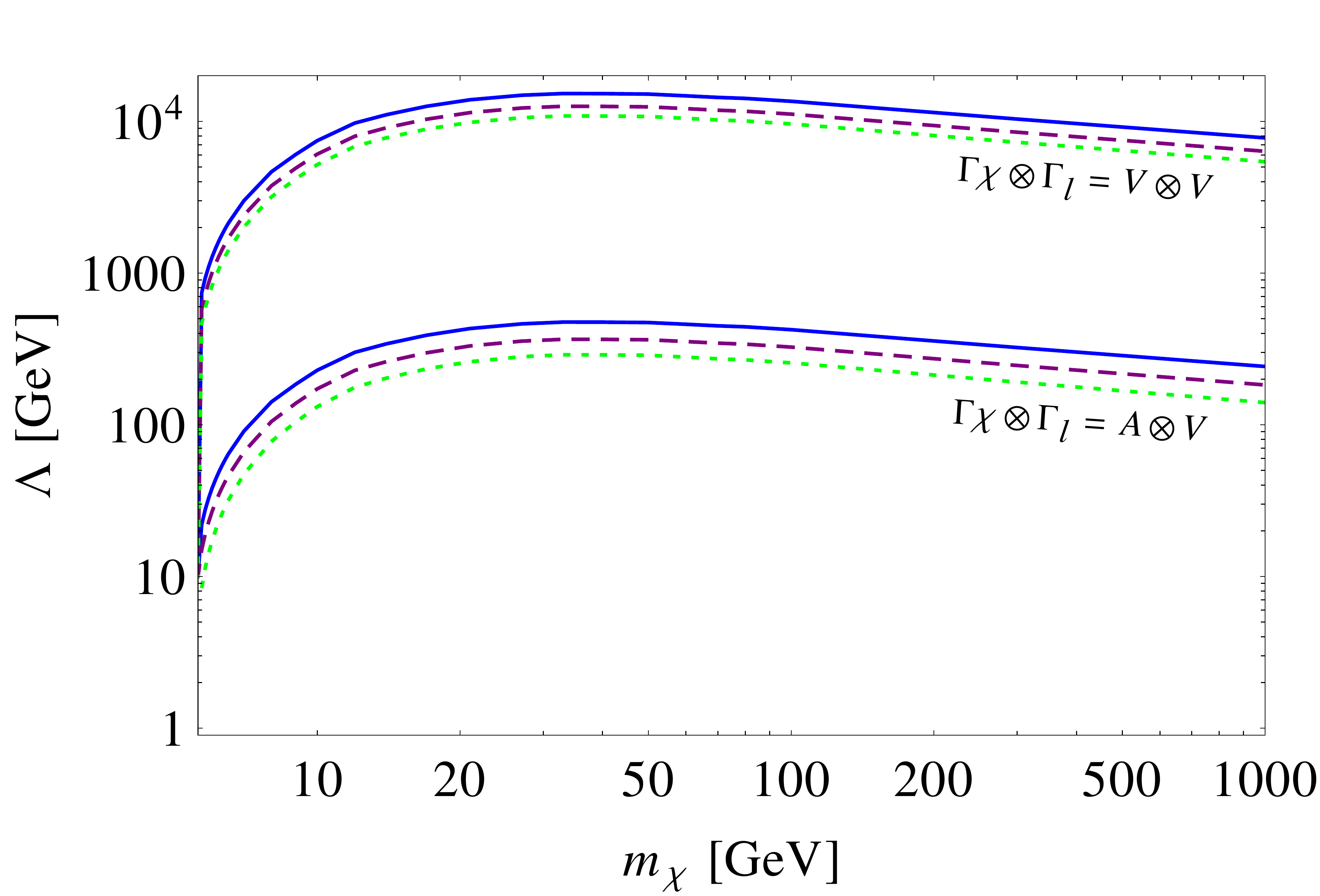}
  \caption{Lower bounds on the cutoff, $\Lambda$, for fermionic DM
    coupling to leptons via a four-fermion effective contact operator,
    using exclusion limits from LUX. Limits are shown for electrons
    (solid, blue), muons (dashed, purple) and taus (dotted, green) for
    both $\Gamma_\chi\otimes\Gamma_\ell = A\otimes V$ and $V\otimes
    V$. We see that the cutoff required for $V\otimes V$ is much
    higher than that for $A\otimes V$.}
\label{Fig:cutoffs}
\end{figure}

To compare these cross sections with results from a given direct
detection experiment, we divide Eq. \ref{eq:d1}, \ref{eq:d2} by the
squared atomic mass number $A^2$ to obtain the cross section per
nucleon.  The most stringent DD results are currently provided by the
LUX experiment~\cite{lux:2014}, for which $A$ = 121 (Xenon).  In
Fig. \ref{Fig:cutoffs} we plot the constraints from the LUX
experiment, in the $m_{\chi}$-$\Lambda$ plane, for both
$\Gamma_\chi\otimes\Gamma_\ell=V\otimes V$ and $A\otimes V$.

\begin{figure}[h]
  \centering
  \includegraphics[width=\columnwidth]{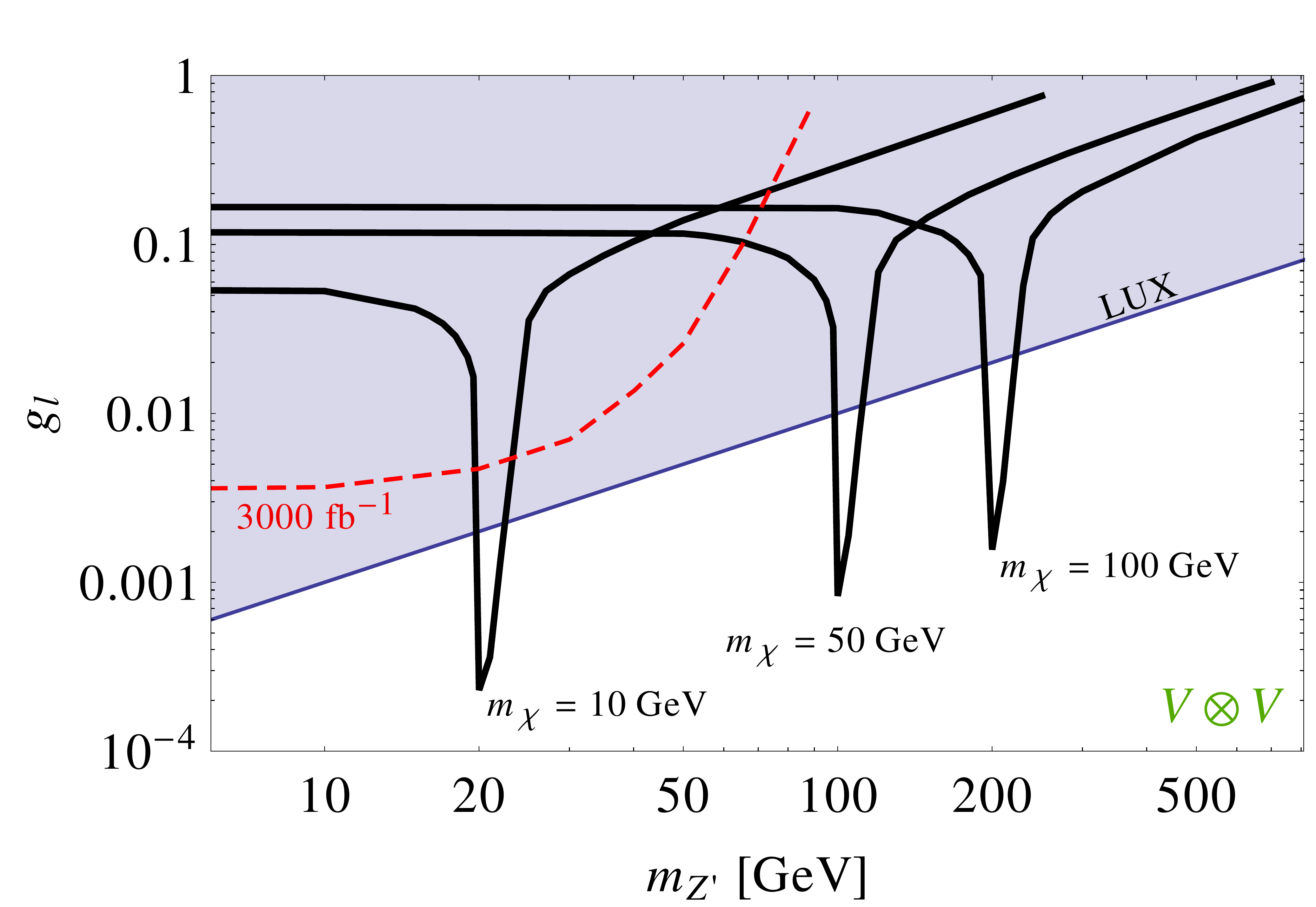}
  \caption{Constraints on the $g_\ell=g_\chi$ and $m_{Z'}$ parameter
    space when $\Gamma_\chi\otimes\Gamma_\ell=V\otimes V$.  The relic
    density curves are shown for $m_{\chi}=10$, 50, and 100 GeV
    (black, solid), while the shaded region is excluded by the LUX
    direct detection results (for $m_\chi = 20 - 100$ GeV).  Also
    shown is the approximate future LHC reach at $\sqrt{s}=14$ TeV
    (see section \ref{sec:lhc}) for coupling to $e$ or $\mu$.  It is
    clear that, for a $\Gamma_\chi\otimes\Gamma_\ell=V\otimes V$
    operator, the current direct detection results already exclude
    most of the parameter space for which the correct relic density is
    obtained.}
\label{Fig:VV}
\end{figure}

As seen in Fig.~\ref{Fig:cutoffs}, the direct detection bounds are
considerably more stringent for the
$\Gamma_\chi\otimes\Gamma_\ell=V\otimes V$ case than for
$\Gamma_\chi\otimes\Gamma_\ell=A\otimes V$.  This is because the DD
cross section for the later is suppressed by a factor of $v_\chi^2$.
The DD and relic density constraints are compared in Fig. \ref{Fig:VV}
for $\Gamma_\chi\otimes\Gamma_\ell=V\otimes V$ and in
Fig.~\ref{fig:mueall22}, \ref{fig:mueall2}, \ref{fig:mueall} for $\Gamma_\chi\otimes\Gamma_\ell=A\otimes
V$.  For $V\otimes V$, the DD bound excludes all the parameter space
for which the relic density constraint is satisfied, except if there
is a very strong resonant enhancement of the annihilation cross
section at $m_\chi=2m_{Z'}$.  In fact, the DD bounds on the $V\otimes
V$ case are much stronger than collider bounds and other constraints.
In comparison, we see that more parameter space is open for $A\otimes
V$, however, even in this case the DD experiments place non-trivial
constraints.

\begin{figure}[h]
\begin{fmffile}{loopc}
\begin{fmfgraph*}(220,55)
\fmfset{arrow_len}{8}
\fmfkeep{loopc}
\fmfleft{i1,i2}
\fmfright{o1,o2}
\fmf{phantom,tension=.5}{i1,v1,i2}
\fmf{phantom,tension=.5}{o1,v4,o2}
\fmf{photon,label=$Z'$}{v1,v2}
\fmf{photon,label=$\gamma$}{v3,v4}
\fmf{fermion,right,tension=.5,label=$\ell$}{v2,v3,v2}
\end{fmfgraph*}
\end{fmffile}
\caption{Kinetic mixing of the $Z'$ and SM photon.}
\label{fig:loope}
\end{figure}
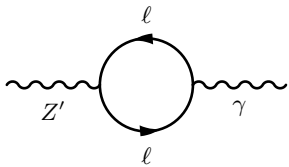

In calculating the direct detection bounds, we have assumed the
DM-nucleus cross section arises entirely from the loop diagram of
Fig.~\ref{fig:loopa}.  However, kinetic mixing of the $Z'$ with the SM
hypercharge $Z'$ gauge boson (the $\epsilon$ term in
Eq. \ref{eq:lagr1}) would also make a contribution to the direct
detection cross section.  In fact, such a term should be present in
the Lagrangian as it will be generated by loop diagrams such as that
in Fig.~\ref{fig:loope}.  This loop contribution to the kinetic mixing
may be estimated as \cite{Holdom:1986}
\begin{equation}
\epsilon=\frac{g_Y g_\ell}{16\pi^2}\textrm{log}\left(\frac{\mu^2}{m_\ell^2}\right),
\label{eq:eps}
\end{equation}
where $\mu$ is a renormalization scale, and clearly leads to the same
DD cross sections as those in Eq.~\ref{eq:d1} and Eq.~\ref{eq:d2}.
Using Eq.~\ref{eq:eps}, we may translate our DD bounds into a
constraint on the kinetic mixing parameter of
$\epsilon\lesssim\mathcal{O}(10^{-3})-\mathcal{O}(10^{-4})$.  These
direct detection bounds on $\epsilon$ are more stringent than those
arising from precision electroweak measurements, which require only
$\epsilon\lesssim\mathcal{O}(10^{-2})$~\cite{Babu:1998, Hook:2010}.

Note that our DD cross sections should actually be considered
naturalness estimates, in the case that the kinetic mixing is
generated by the loop diagrams alone.  More generally, $\epsilon$ is a
free parameter which runs with scale.  In principle, the direct
detection signal could be suppressed if $\epsilon$ happened to run to
a very small value at low scale.  We regard this as unnatural, as it
would involve a fine-tuned cancellation of the contribution of the
loop diagram with a bare term in the Lagrangian.  However, it is
important to appreciate that the DD limits are not watertight
constraints, but simply naturalness bounds.  

\section{Constraints from $(g-2)_\ell$, LEP and other searches}
\label{sec:constraints}

We now outline further constraints on our scenario.  Unlike the DD
bounds, these additional limits depend sensitively on the lepton
flavour.  Measurements of $(g-2)_\ell$ constrain the coupling of a
$Z'$ to each of the lepton flavours, resulting in a strong bound for
the $\mu$ flavour, and weaker bounds for $e$ and $\tau$.  Very
stringent $Z'$ bounds from LEP apply to the electron flavour alone, as
do the LEP mono-photon bounds.

\medskip

\noindent
$\bm{(g-2)_\ell}$: A $Z'$ which couples to leptons will make a
contribution to the lepton anomalous magnetic dipole of \cite{fayet:2007}
\begin{equation}
\Delta(g-2)_\ell \sim \frac{g_\ell^2}{6\pi^2}\frac{m_\ell^2}{m_{Z'}^2}.
\end{equation}

Upper limits on any additional contribution to $(g-2)_\ell$ are
$4\times10^{-10}, 8\times10^{-9}$ and $8\times10^{-2}$ for the electron, muon and tau
respectively \cite{pdg:2013}. This requires the $Z'$ coupling strengths to be
\begin{subequations}
 \begin{equation}
g_e\lesssim0.3\frac{m_{Z'}}{\textrm{GeV}},
 \end{equation}
 \begin{equation} 
g_\mu\lesssim6\times10^{-3}\frac{m_{Z'}}{\textrm{GeV}},
 \end{equation}
 \begin{equation}
g_\tau\lesssim\frac{m_{Z'}}{\textrm{GeV}}.
 \end{equation}
\end{subequations}

\medskip

\noindent\textbf{Neutrino scattering}: The Liquid Scintillator Neutrino Detector (LSND) measured the cross section for the elastic scattering process $\nu_e+e^-\rightarrow\nu_e+e^-$, placing a further constraint on the $Z'$ coupling strength to electrons \cite{LSND:2001},
 \begin{equation}
g_e\lesssim3\times10^{-3}\frac{m_{Z'}}{\textrm{GeV}}.
 \end{equation}
This constraint is comparable to the LUX direct detection bound on $g_e$ for the $A\otimes V$ case. 

\medskip

\noindent\textbf{LEP-II Z' constraints}: 
The coupling of a $Z'$ to electrons is constrained by results of the
LEP-II experiments.  For $Z'$ masses greater than 209 GeV, the largest
center-of-mass energy at which LEP operated, the constraints are
expressed in terms of four-fermion contact operators, known as the
compositeness bounds~\cite{LEP:2003}.  For a vector coupling to
electrons this bound can be expressed as~\cite{Buckley:2011,LEP:2003}
\begin{equation}
g_e\lesssim0.044\times m_{Z'}/(200 \textrm{ GeV}).
 \end{equation}
For $Z'$ masses below about 200 GeV, the four-fermion description is not
valid, as the $Z'$ mass is not large compared to the LEP beam energy,
and resonant production is possible.  A conservative limit can be
taken as $g \alt 0.04$ for $m_{Z'} \lesssim 200$~GeV.  Much stronger limits
should hold for a $Z'$ mass close to one of the centre of mass
energies at which LEP ran, however no detailed analysis exists.

\medskip

\noindent\textbf{LEP-II mono-photon constraints}:
Monophoton searches at LEP-II place bounds on the
couplings~\cite{Fox:2011}, which again are relevant only when the $Z'$
couples to electrons.  These constraints depend sensitively on the
$Z'$ decay width and thus on the ratio $g_e/g_\chi$.  If we assume
$g_e \simeq g_\chi$, then for $m_\chi\lesssim m_{Z'}/2$ and
$m_{Z'}\gtrsim 30$~GeV these constraints are stronger than LUX, but
are comparable to the LEP $Z'$ bounds.  For masses outside of this
range, LUX is more constraining.  

\medskip

\noindent
\textbf{Electroweak Precision Measurements}: In addition to the limits on kinetic mixing as discussed in section II, there are constraints on the ratio of the decay width of a $Z'$ which couples to electrons and $m_{Z'}$ from precision measurements of the line shape of the $Z^0$ \cite{Babu:1998, Hook:2010}. However they do not constrain the $g_e-m_{Z'}$ parameter space any further than the limits listed above.

\section{LHC Phenomenology}
\label{sec:lhc}

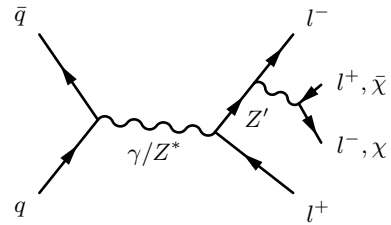
\begin{figure}[H]
\centering
\begin{fmffile}{diagramo}
\begin{fmfgraph*}(120,60)
\fmfset{arrow_len}{8}
\fmfleft{i1,i2}
\fmfright{o1,o2,o3,o4}
\fmflabel{$q$}{i1}
\fmflabel{$\bar{q}$}{i2}
\fmflabel{$l^+$}{o1}
\fmflabel{$l^-,\chi$}{o2}
\fmflabel{$l^+,\bar{\chi}$}{o3}
\fmflabel{$l^-$}{o4}
\fmf{fermion}{i1,v1,i2}
\fmf{boson,label=$\gamma/Z^*$}{v1,v2}
\fmf{fermion,tension=1}{o1,v2,v3,o4}
\fmffreeze
\fmf{boson,label=$Z'$}{v3,v4}
\fmf{fermion,tension=1}{o3,v4,o2}
\end{fmfgraph*}
\end{fmffile}
\vspace{4 mm}
\caption{
Production of the $Z'$ at a hadron collider.  The $Z'$ is radiated from a lepton in the Drell Yan process, and subsequently decays to either leptons or DM.}
\label{Fig:decay} 
\end{figure}

\begin{figure}[h]
  \centering
  \includegraphics[width=\columnwidth]{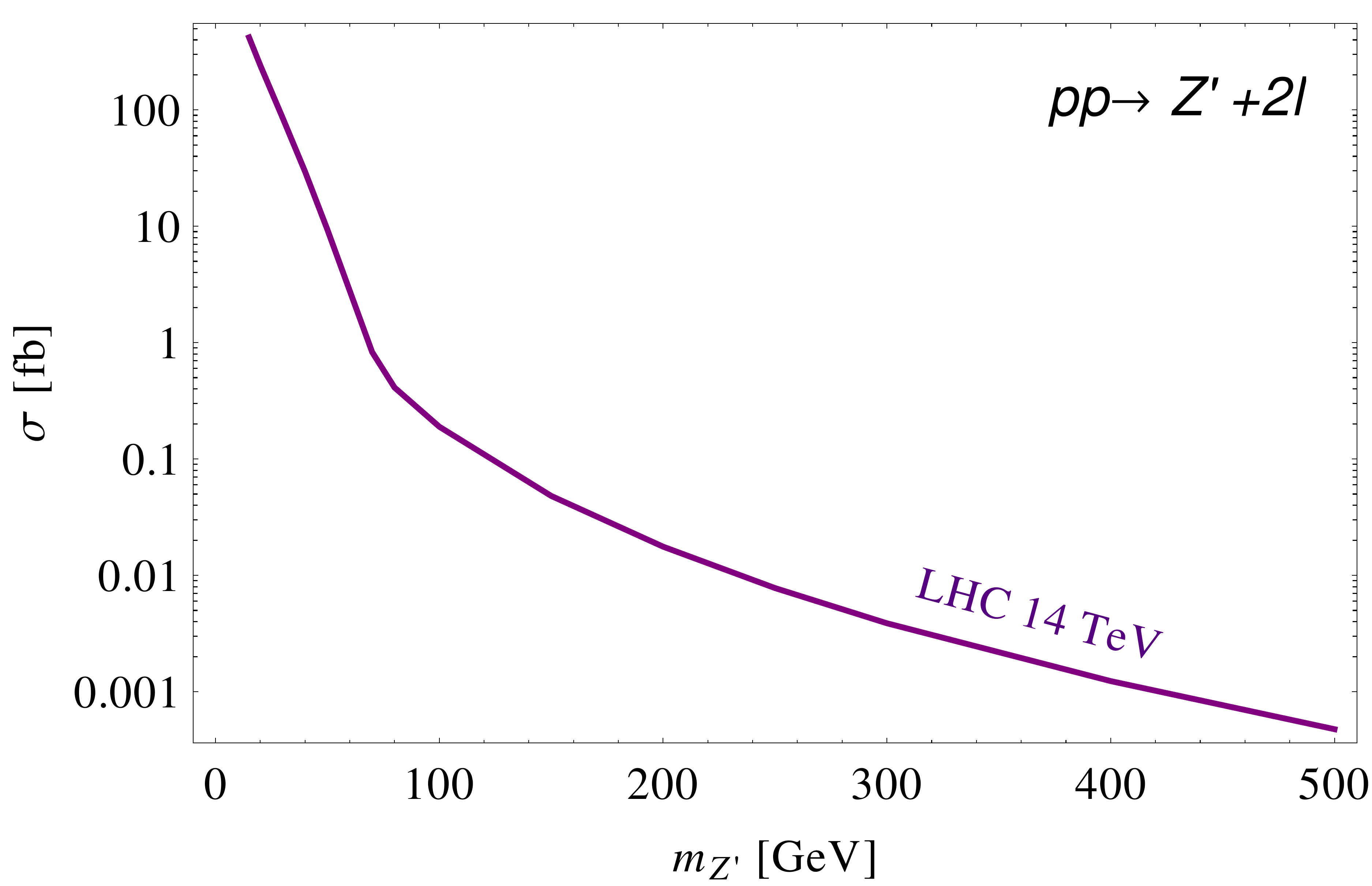}
  \caption{$Z'$ production cross section at the 14 TeV LHC, via the process $pp\rightarrow \ell^+ \ell^- Z' $. We have set $g_\ell = 0.1$.}
\label{Fig:zpcrosssec}
\end{figure}

We now consider the LHC phenomenology for a leptophilic $Z'$.
Because the $Z'$ does not couple directly to quarks, the lowest order
$Z'$ production process is $pp \rightarrow \ell^+ \ell^- Z'$, in which
a $Z'$ is radiated from a lepton in a Drell-Yan process, as shown
Fig. \ref{Fig:decay}.  
The $Z'$ production cross section is shown in
Fig. \ref{Fig:zpcrosssec}.  The cross section is large when $m_{Z'}
< m_Z$, because the process in Fig. \ref{Fig:decay} can proceed via
an on-shell $Z$.  For $m_{Z'} > m_Z$, however, the cross section falls
rapidly with increasing $Z'$ mass, such that detecting a $Z'$ with
mass beyond about $500$~GeV would be challenging.

The $Z'$ decays either to DM (or neutrinos)
or to charged leptons, resulting in a pair of opposite sign di-leptons
plus missing $E_T$, or two pairs of opposite sign di-leptons,
respectively.  The 2-lepton plus missing $E_T$ signal competes with
substantial SM backgrounds, in particular from the process $Z$+jets,
such that detection prospects are poor.  However, the 4-lepton signal
is very distinctive and is examined in detail below.  

\vspace{2 mm}
\hspace{12 mm}

The signal rates depend on the branching fraction of the $Z'$ to charged
leptons or DM, which depend on the mass and coupling strength of the $Z'$.  
For $g_\chi=g_\ell$, and in the limit $m_{Z'}\gg m_\chi$, we have 
 \begin{equation}
Br(Z'\rightarrow \ell\ell)=Br(Z'\rightarrow \chi\chi)=2 Br(Z'\rightarrow \nu_\ell\nu_\ell).
\label{eq:branching}
 \end{equation}
For other parameters, the branching ratios are evaluated numerically in our analysis.

\subsection{$Z'$ Decay to Leptons}

We now consider the 4 lepton process, $pp\rightarrow \ell^+\ell^-Z'
\rightarrow \ell^+\ell^-\ell^+\ell^-$, in detail.  The main SM
backgrounds for this process are
\begin{subequations}
 \begin{equation}
pp \rightarrow \ell^+\ell^- Z \rightarrow\ell^+\ell^-\ell^+\ell^-,
\label{eq:lepback}
 \end{equation}
 \begin{equation} 
pp \rightarrow ZZ \rightarrow \ell^+\ell^-\ell^+\ell^-,
 \end{equation}
\end{subequations}
with Eq.~\ref{eq:lepback} making the dominant contribution.  The
$pp\rightarrow 4\ell$ cross section has been measured at the $Z$
resonance by the LHC experiments, ATLAS and CMS.  The most
constraining limits arise from the ATLAS analysis, which used an
$\sqrt{s}=8$ TeV dataset at an integrated luminosity of 20.7
$fb^{-1}$, and measured the number of events to be consistent with the
SM expectation.  A similar analysis performed by CMS used only the
$\sqrt{s}=7$ TeV data \cite{CMS:2013} and is less constraining.  We
reproduce the ATLAS analysis to find the current exclusion limits for
our $Z'$ model using the $Z\rightarrow4l$ search
\cite{ATLAS:2014}.\footnote{A similar analysis has been performed
  recently for the 4 muon final state in
  Refs. \cite{Alt:2014,Hari:2013}.}  We also simulate events at
$\sqrt{s}=14$ and higher luminosities, to project the future reach of
the LHC.

To simulate our $Z'$ signal and the relevant SM background, we
implement our model with FeynRules \cite{Feyn:2009}, generate parton
level events in MadGraph \cite{Madg:2011} and then interface with
Pythia \cite{Pythia:2008} to produce hadronic level events.  For
processes involving electrons, we also interface the Pythia output
with the PGS detector simulation \cite{PGS}.  Finally, we use MadAnalysis
\cite{Mada:2013} to analyse the events.
We determine the significance according to
\begin{equation}
\sigma=N_{Z'}/\sqrt{N_{Z'}+N_{SM}},
\end{equation}
where $N_{Z'}$ is the number of simulated events for the $Z'$ model,
and $N_{SM}$ is the number of ATLAS events observed, which is
consistent with the predicted number of SM events.  Excluded
parameters are those which have a deviation from the SM of
$\sigma\gtrsim3.0$.  We neglect systematic uncertainties, as they are
very low for our purely leptonic final states \cite{ATLAS:2014}.
Given that the number of signal and background events are comparable, a
small systematic uncertainty will not have a significant effect on the
results.

We consider only the $Z\rightarrow4\mu$ and $Z\rightarrow4e$ part of
the ATLAS analysis.  Mixed flavour final states are not possible,
because we assume the $Z'$ couples to a single lepton flavour.  For
the case of the $4\mu$ signal, we perform different analyses for low
mass ($m_{Z'}<m_Z$) and high mass ($m_{Z'}>m_Z$) $Z'$ bosons.  For the
$4e$ signal, however, we perform only the low mass analysis, as the
LEP $Z'$ searches already eliminate the high mass parameter space
that could be probed at the LHC.

There is no available analysis for a four tau final state, as tau
reconstruction is significantly more difficult and suffers a much
lower efficiency than detecting $\mu$ or $e$.  Therefore, there are no
current collider constraints which can be placed on a scenario in
which the $Z'$ couples only to the tau flavour.

\subsubsection{Four Electron Final State: $m_{Z'}<m_Z$}

We replicate the ATLAS $pp\rightarrow 4e$ analysis at the $Z$
resonance, for which the candidate events have two pairs of opposite
sign electrons.  The following selection cuts are made:

\begin{itemize}
  \item $p_{T,e} > 7$ GeV and  $|\eta|<2.47$ for individual electrons
  \item Candidate separation of $\Delta R_{ee}>0.1$ 
  \item $M_{e^-,e^+} >$ 20, 5 GeV for the leading pair and next to leading pair in momentum
  \item $p_{T,e} >$ 20, 15, 10 GeV for the leading three electrons
  \item Invariant mass of electron quadruplet is restricted to events near the $Z$ resonance: $80<M_{4e}<100$ GeV
\end{itemize}

For the four electron case, the hadronic level events generated by
Pythia \cite{Pythia:2008} are interfaced with the PGS detector
simulator before being interfaced to MadAnalysis at reconstruction
level \cite{Mada:2013}.  This is necessary for electrons, as they are
reconstructed from energy clusters in an electromagnetic calorimeter,
which are matched to reconstructed electron tracks in the inner
detector \cite{ATLAS:2012}, for which detector effects and
efficiencies are not negligible.  With this procedure, our simulated
number of SM events was consistent with that measured by ATLAS.

The current ATLAS exclusion based on the $4e$ process at 20.7
$fb^{-1}$ is shown in Fig. \ref{fig:mueall}, assuming $g_\ell =
g_\chi$.  We also show future discovery curves at the higher
luminosities of 300 and 3000 $fb^{-1}$.

\begin{figure*}[t]
\centering
\begin{subfigure}
\centering
  \includegraphics[width=\columnwidth]{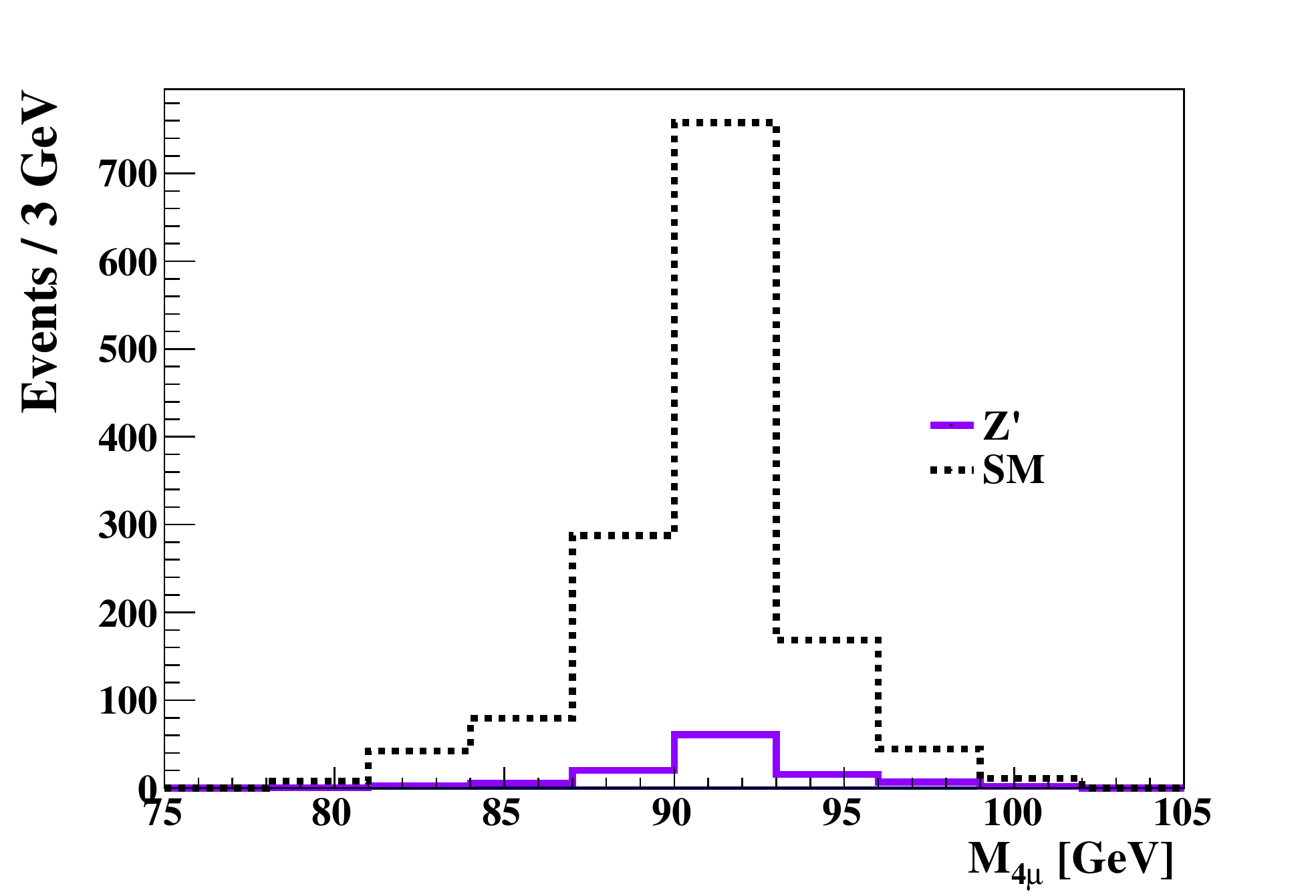}
\end{subfigure}\hfill
\begin{subfigure}
\centering
  \includegraphics[width=\columnwidth]{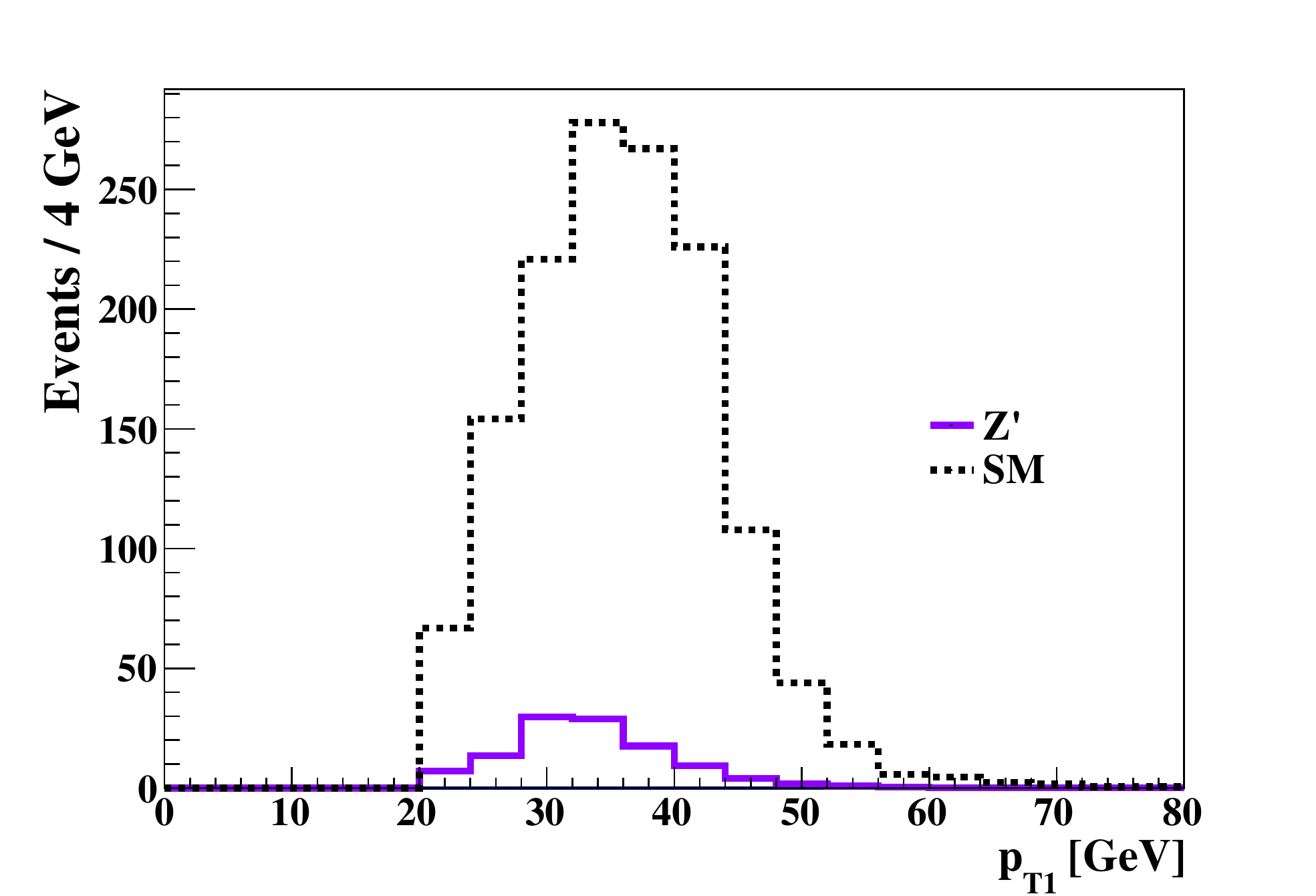}
\end{subfigure}\hfill
\caption{Invariant mass for four muons (left) and transverse momentum $p_T$ for leading in $p_T$ muon (right) for $pp\rightarrow4\mu$ in the SM and $Z'$ model (with $m_{Z'}=60$ GeV, $m_{\chi}=10$ GeV, $g_\mu=g_{\chi}=0.1$), at $\sqrt{s}=14$ TeV and $\mathcal{L}=300$ $fb^{-1}$. The peak in the four muon invariant mass spectrum is a reconstruction of the $Z$ mass.}
\label{fig:mus}
\end{figure*}

\begin{figure*}[t]
\centering
\begin{subfigure}
\centering
  \includegraphics[width=\columnwidth]{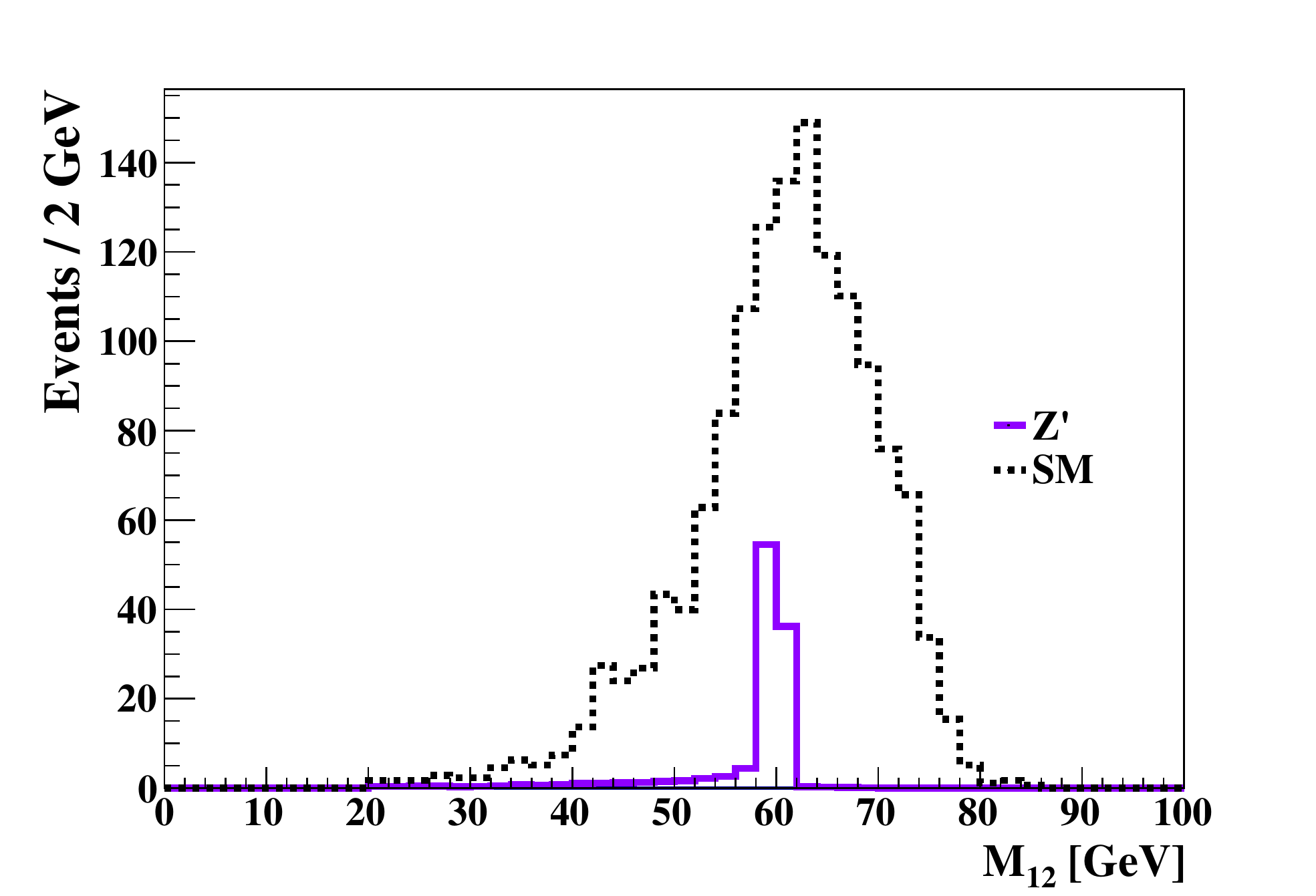}
\end{subfigure}\hfill
\begin{subfigure}
\centering
  \includegraphics[width=\columnwidth]{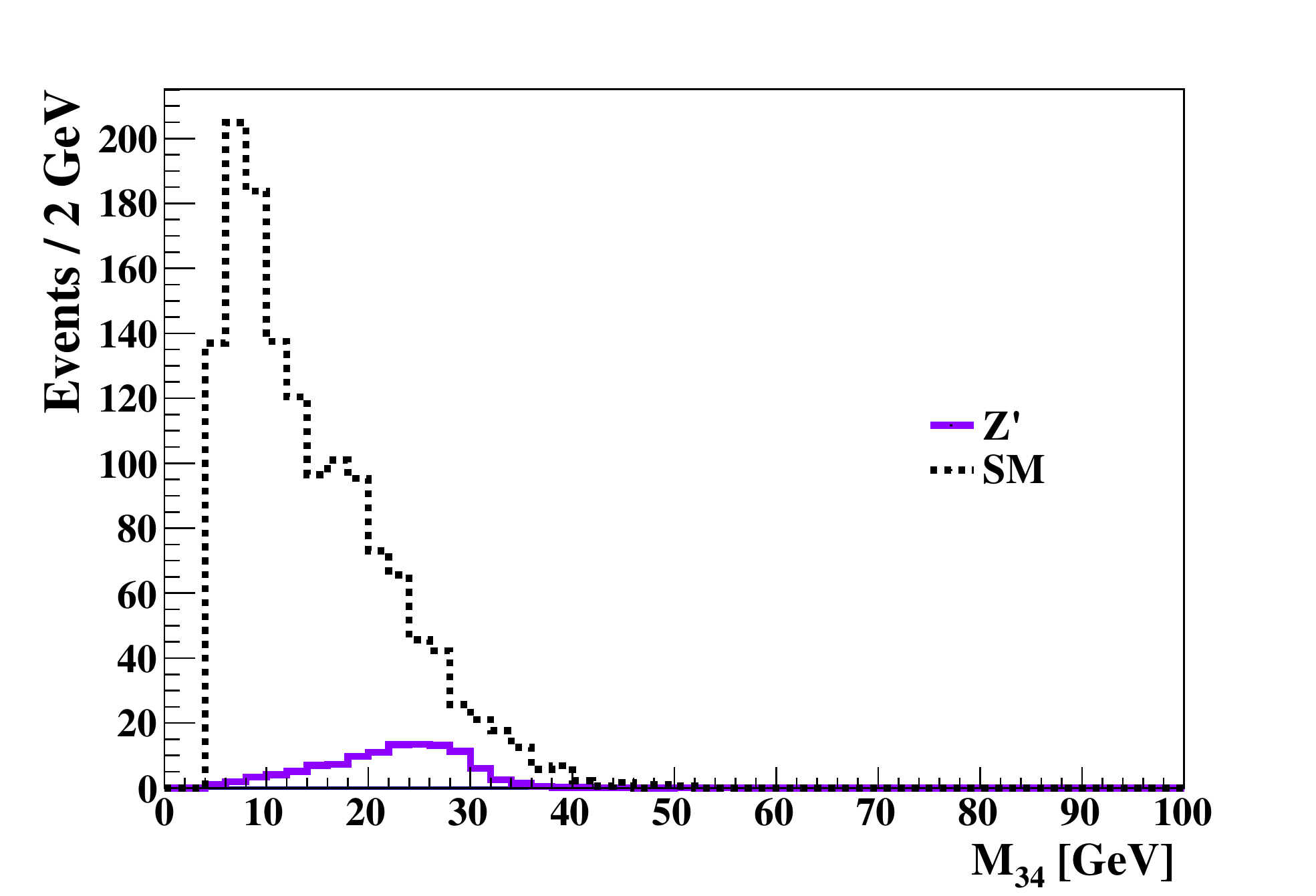}
\end{subfigure}\hfill
\caption{Invariant mass for first and second leading muons in $p_T$ (left) and third and fourth leading muons in $p_T$ (right) for $pp\rightarrow4\mu$ in the SM and $Z'$ model (with $m_{Z'}=60$ GeV, $m_{\chi}=10$ GeV, $g_\mu=g_{\chi}=0.1$), at $\sqrt{s}=14$ TeV and $\mathcal{L}=300$ $fb^{-1}$. The mass of the $Z'$ can be seen clearly as the resonance at $m_{Z'} = 60$ GeV in the invariant mass spectrum $M_{12}$.}
\label{fig:mus2}
\end{figure*}

\subsubsection{Four Muon Final State: $m_{Z'}<m_Z$}

We replicate the ATLAS $pp\rightarrow 4\mu$ analysis at the $Z$
resonance, for which the candidate events have two pairs of opposite
sign muons.  The following selection cuts are made:
\begin{itemize}
  \item $p_{T,\mu} > 4$ GeV and $|\eta|<2.7$ for individual muons
  \item Candidate separation of $\Delta R_{\mu\mu}>0.1$ 
  \item $M_{\mu^-,\mu^+} >$ 20, 5 GeV for the leading pair and next to leading pair in momentum
  \item $p_{T,\mu} >$ 20, 15, 8 GeV for the leading three muons
  \item Invariant mass of muon quadruplet is restricted to events near the $Z$ resonance: $80<M_{4\mu}<100$ GeV
\end{itemize}

We do not perform a detector simulation for the $pp\rightarrow4\mu$
analysis, as detection efficiencies are very high and the small
smearing of data due to detector effects has a negligible effect on our
results.

\begin{figure*}[t]
\centering
\begin{subfigure}
\centering
  \includegraphics[width=\columnwidth]{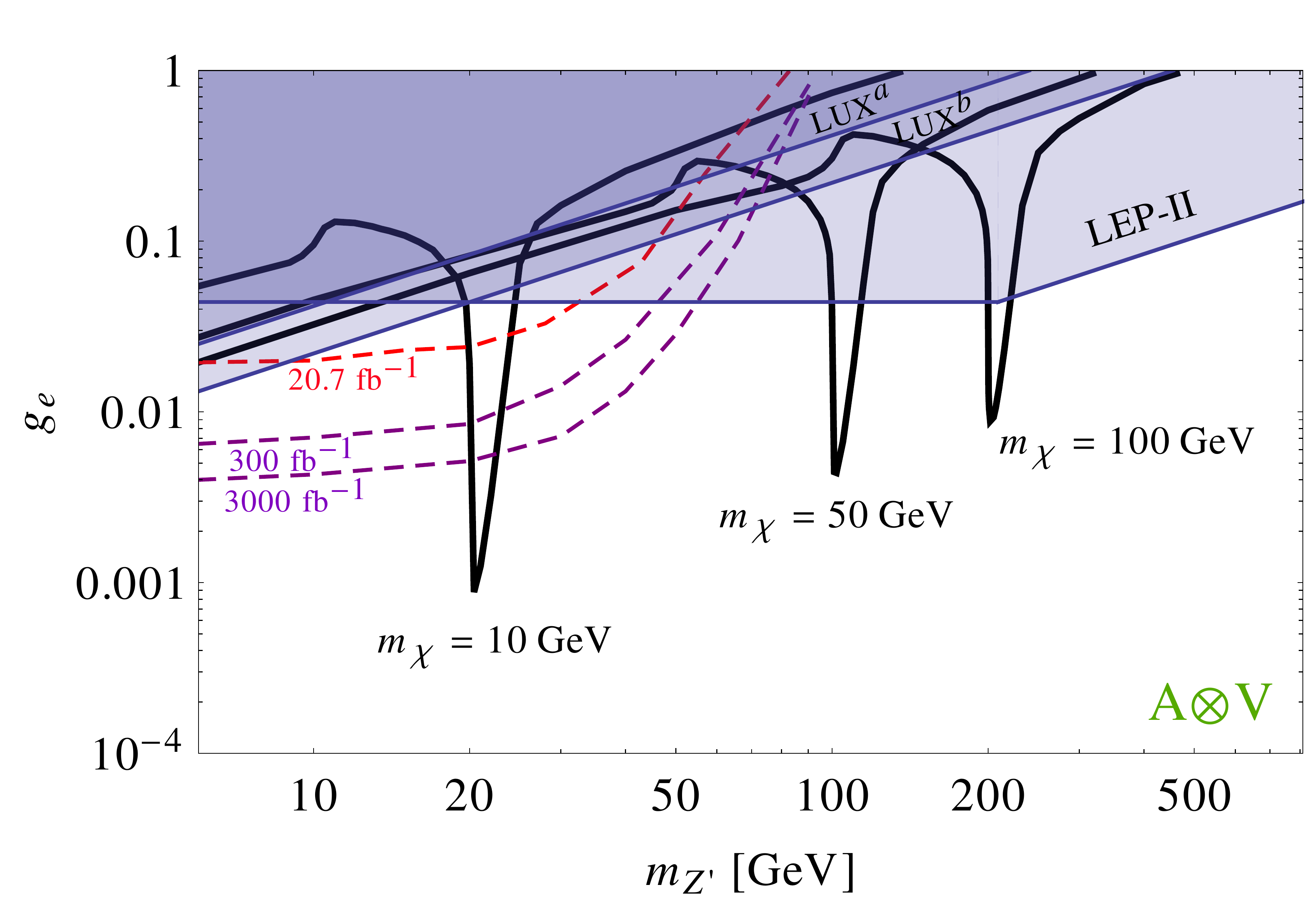}
\end{subfigure}\hfill
\begin{subfigure}
\centering
  \includegraphics[width=\columnwidth]{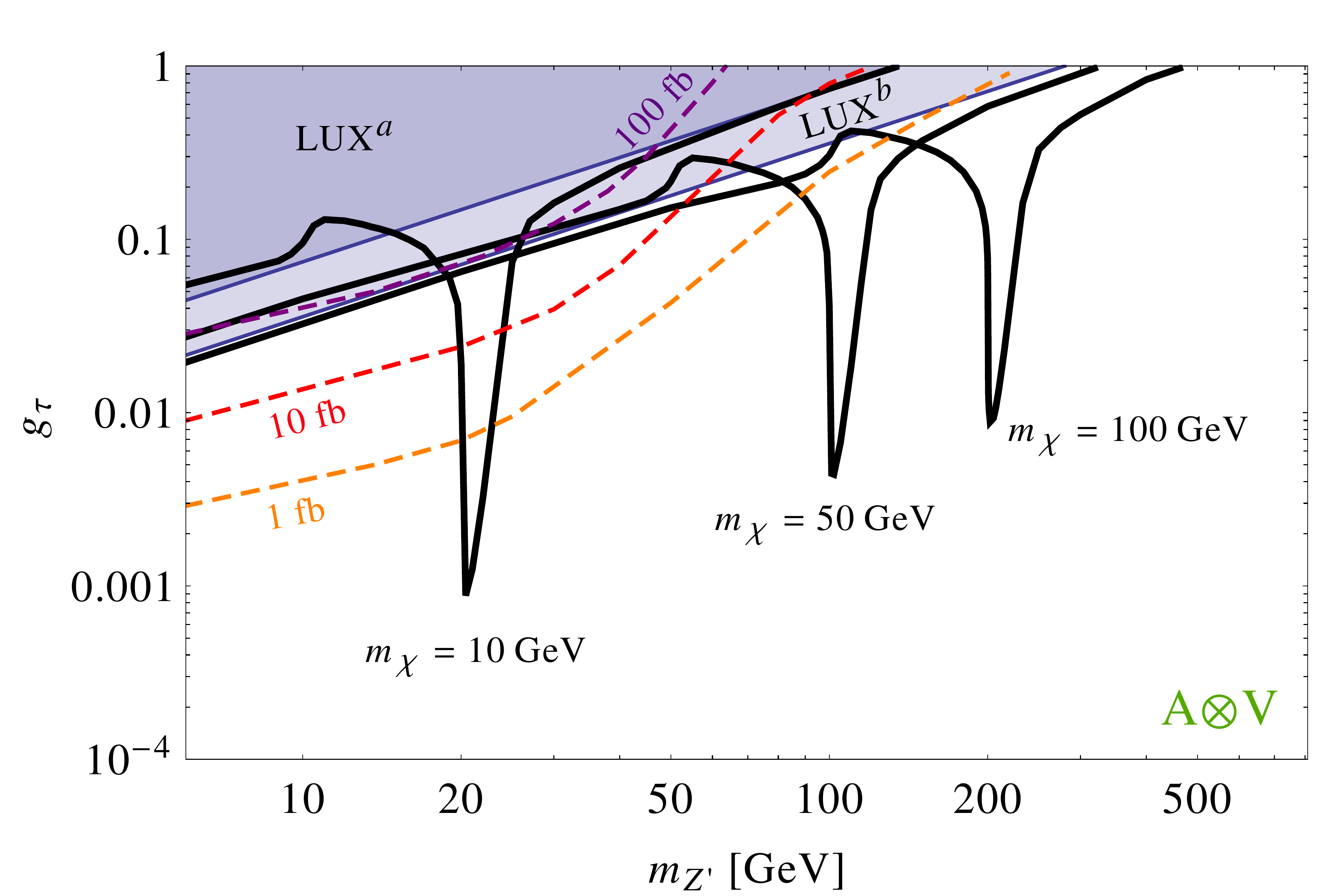}
\end{subfigure}\hfill
\caption{Parameter space for $g_e$ vs. $m_{Z'}$ with $g_\chi=g_e$
  (left) and $g_\tau$ vs. $m_{Z'}$, with $g_\chi=g_\tau$ (right).  Two
  excluded regions from LUX direct detection are shown corresponding
  to two different DM mass ranges: LUX$^a$ is for approximately
  $m_\chi=10,1000$ GeV and LUX$^b$ is for the range $m_\chi=20-100$
  GeV.  Relic density curves are shown in black for $m_\chi=10,50,100$
  GeV.  Exclusions from LEP-II (electrons only) shown on the left.
  Dashed lines on the electron plot (left) are ATLAS exclusions and
  reaches: top is ruled out by ATLAS data at $\sqrt{s}=8$ TeV and
  $\mathcal{L}=20.7$$fb^{-1}$, middle is the ATLAS discovery reach at
  $\sqrt{s}=14$ TeV and $\mathcal{L}=300$$fb^{-1}$, and bottom is the
  ATLAS discovery reach at $\sqrt{s}=14$ TeV and $\mathcal{L}=3000$
  $fb^{-1}$. The LHC limits all assume $m_\chi=10$~GeV.  For the tau
  plot (right), there are no collider limits due to low efficiencies
  for reconstructing a four tau final state.  Instead, dashed lines
  show contours of $Z'$ production cross sections of 1, 10 and 100
  $fb$.}
\label{fig:mueall}
\end{figure*}

The current ATLAS exclusion based on the $4\mu$ process at 20.7
$fb^{-1}$ is shown in Fig. \ref{fig:mueall}, assuming $g_\ell =
g_\chi$.  We also show future discovery curves at the higher
luminosities of 300 and 3000 $fb^{-1}$.  
We show a selection of kinematic plots in Fig. \ref{fig:mus},
\ref{fig:mus2} for an example choice of parameters: $m_{Z'}=60$ GeV,
$m_{\chi}=10$ GeV, $g_\mu=g_{\chi}=0.1$.  (The relevant $Z'$ branching
fractions are $Br(\mu^{+}\mu^{-}) = 0.428$,
$Br(\bar{\nu}_\mu\nu_{\mu}) = 0.214$ and $Br(\bar{\chi}\chi) =
0.358$.)  These parameters are allowed by the $\sqrt{s}=8$ TeV ATLAS
results, but sit on the $3\sigma$ curve corresponding to
$\mathcal{L}=300$ $fb^{-1}$ at $\sqrt{s}=14$ TeV, and thus can be
discovered or ruled out with future LHC data.

\begin{figure*}[t]
\centering
\begin{subfigure}
\centering
  \includegraphics[width=\columnwidth]{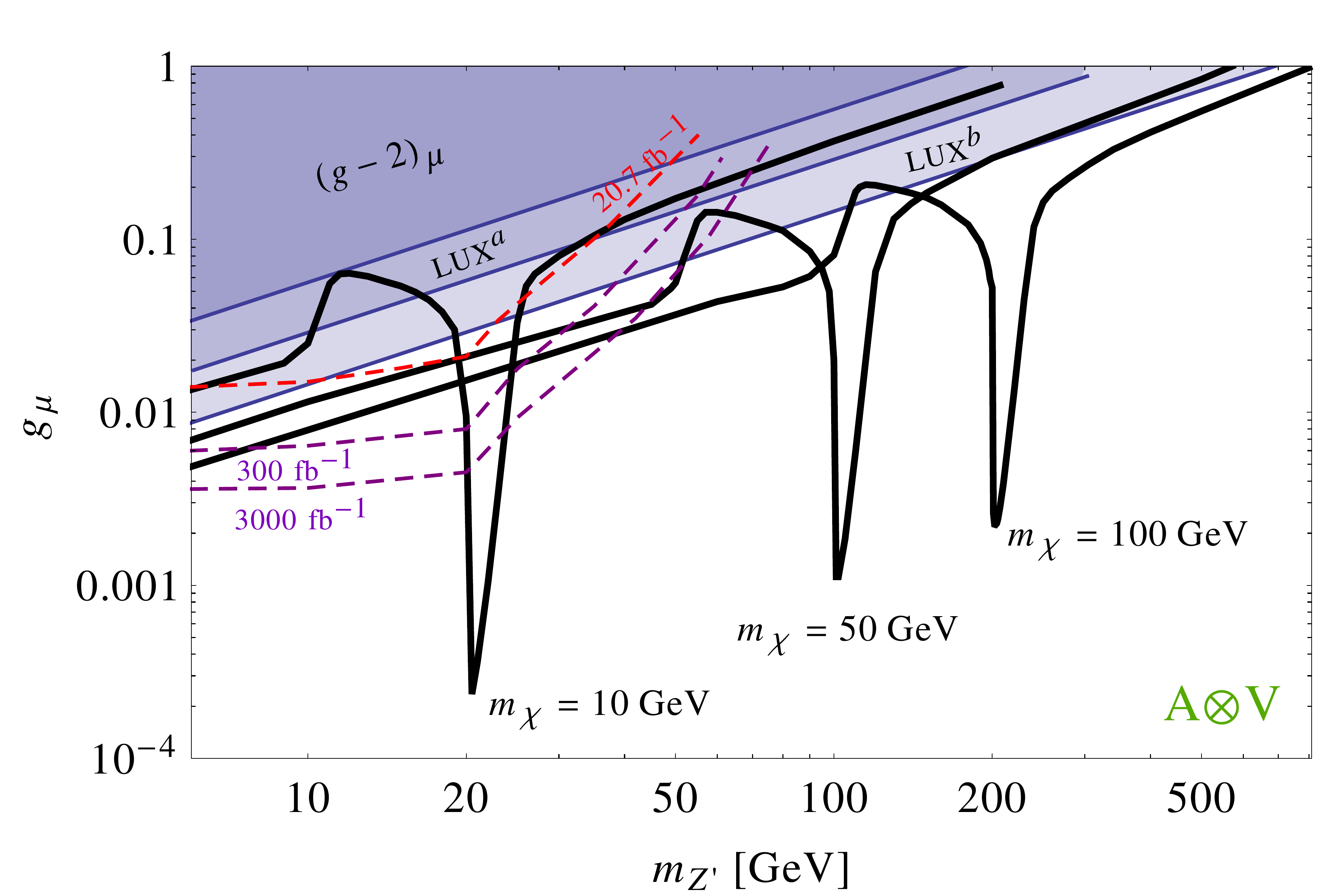}
\end{subfigure}\hfill
\begin{subfigure}
\centering
  \includegraphics[width=\columnwidth]{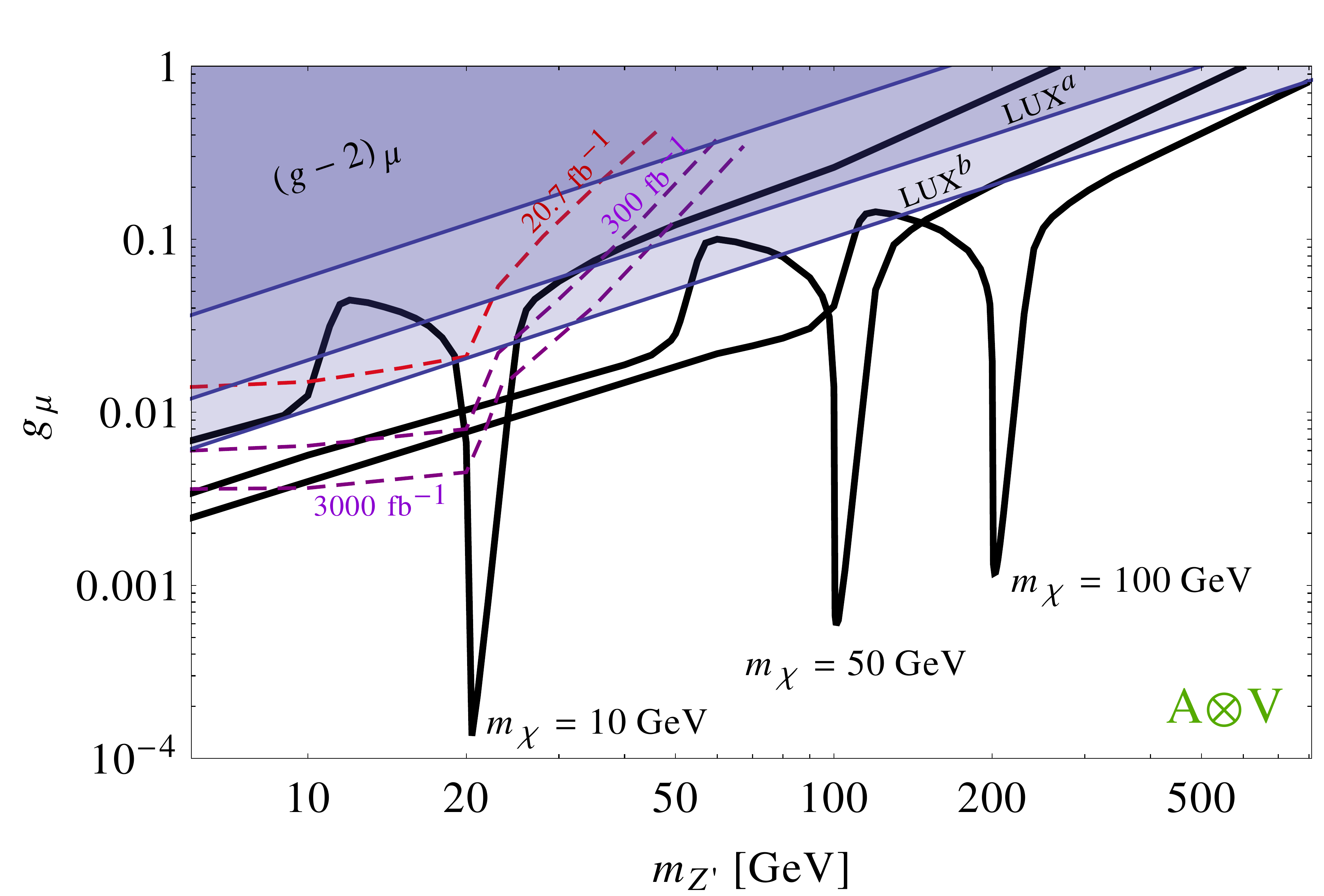}
\end{subfigure}\hfill
\caption{Parameter space for $g_\mu$ vs. $m_{Z'}$, where
  $g_\chi=4g_\mu$ (left) and $g_\chi=8g_\mu$ (right). Shown are
  excluded regions from $(g-2)_\mu$, and two excluded regions from LUX
  direct detection are shown corresponding to two different DM mass
  ranges: LUX$^a$ is for approximately $m_\chi=10,1000$ GeV and
  LUX$^b$ is for the range $m_\chi=20-100$ GeV.  Relic density curves
  are shown in black for $m_\chi=10,50,100$ GeV.  Dashed lines are
  ATLAS exclusions and reaches: top is ruled out by ATLAS data at
  $\sqrt{s}=8$ TeV and $\mathcal{L}=20.7$$fb^{-1}$, middle is the
  ATLAS discovery reach at $\sqrt{s}=14$ TeV and
  $\mathcal{L}=300$$fb^{-1}$, and bottom is the ATLAS discovery reach
  at $\sqrt{s}=14$ TeV and $\mathcal{L}=3000$$fb^{-1}$. These
  exclusion/discovery curves are the same as for the $g_\mu=g_\chi$
  case when $m_{Z'}<2m_\chi$.  The LHC limits all assume
  $m_\chi=10$~GeV.}
\label{fig:mueall2}
\end{figure*}

Note that the choice of $g_\chi$ affects the cross section by
controlling the relative sizes of the $Z'$ branching ratios to lepton
or dark matter final states.  We can weaken constraints from the four
muon search by increasing the $Z'$ coupling strength to dark matter,
with results for $4g_\mu=g_\chi$ and $8g_\mu=g_\chi$ shown in
Fig. \ref{fig:mueall2}.

\begin{figure*}[t]
\centering
\begin{subfigure}
\centering
  \includegraphics[width=\columnwidth]{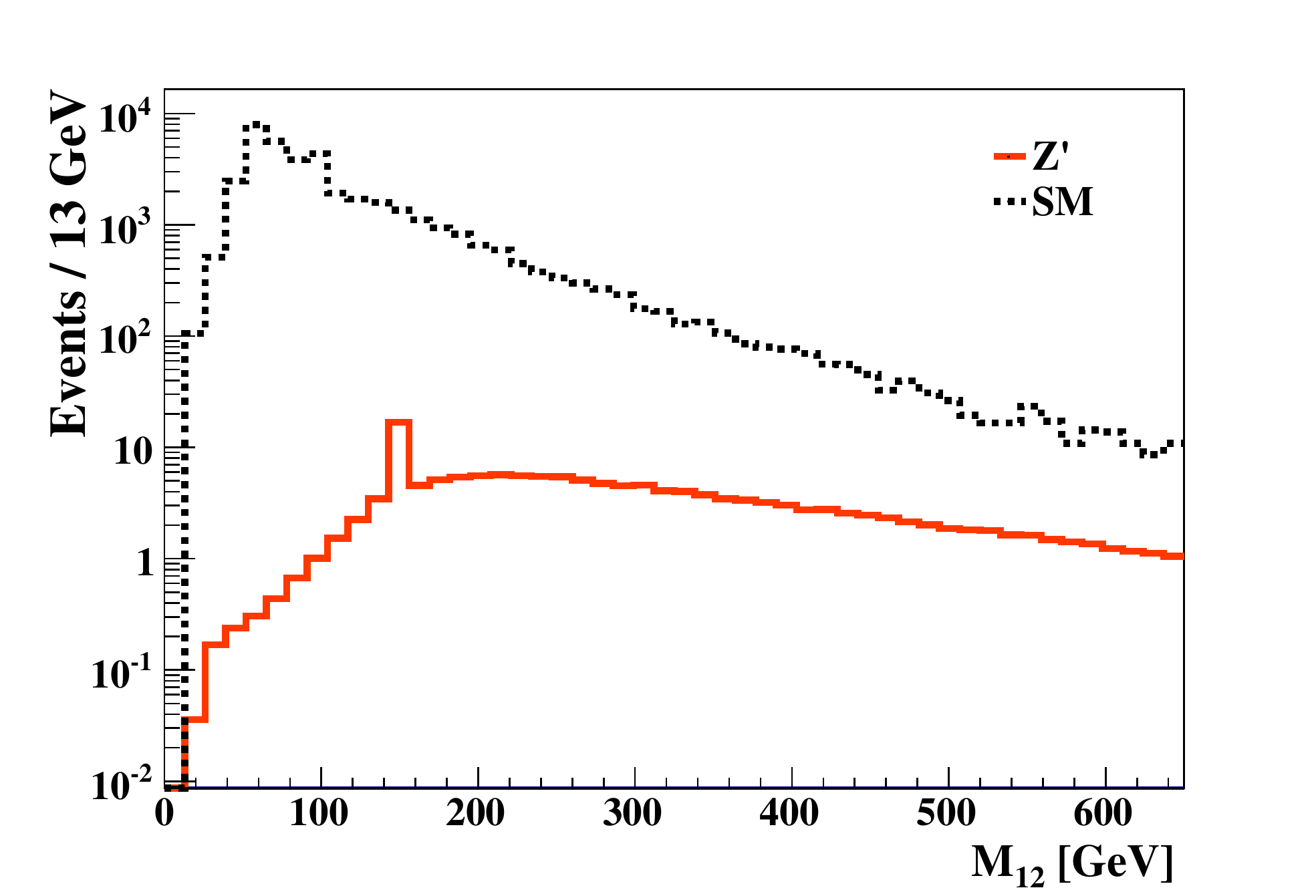}
\end{subfigure}\hfill
\begin{subfigure}
\centering
  \includegraphics[width=\columnwidth]{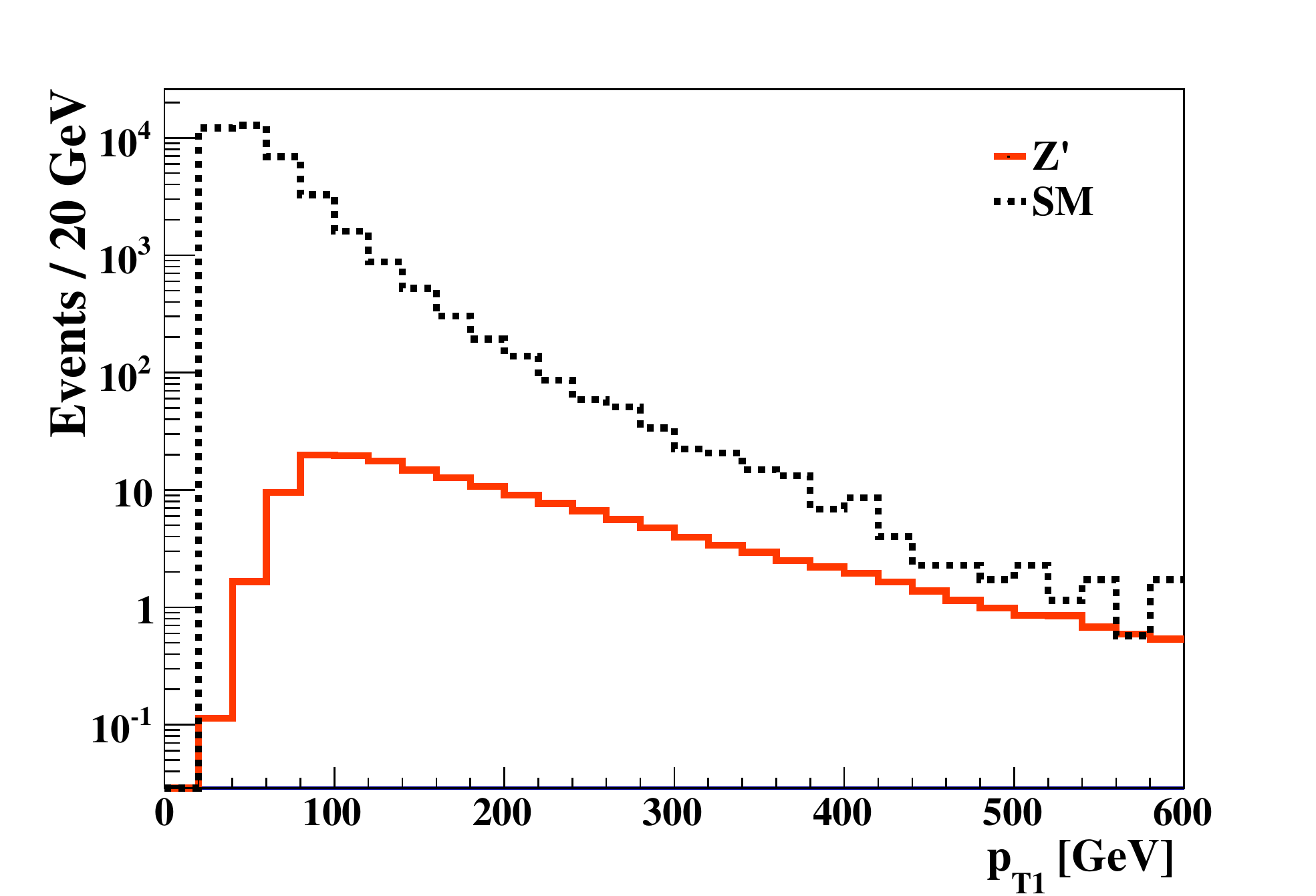}
\end{subfigure}\hfill
\caption{Invariant mass for first and second leading muons in $p_T$ (left) and transverse momentum $p_T$ for $p_T$ leading muon (right) both before cuts, for $pp\rightarrow4\mu$ in the SM and $Z'$ model (with $m_{Z'}=150$ GeV, $m_{\chi}=10$ GeV, $g_\mu=g_{\chi}=0.19$), at $\sqrt{s}=14$ TeV and $\mathcal{L}=3000$ $fb^{-1}$.}
\label{fig:mus3}
\end{figure*}

\begin{figure*}[t]
\centering
\begin{subfigure}
\centering
  \includegraphics[width=\columnwidth]{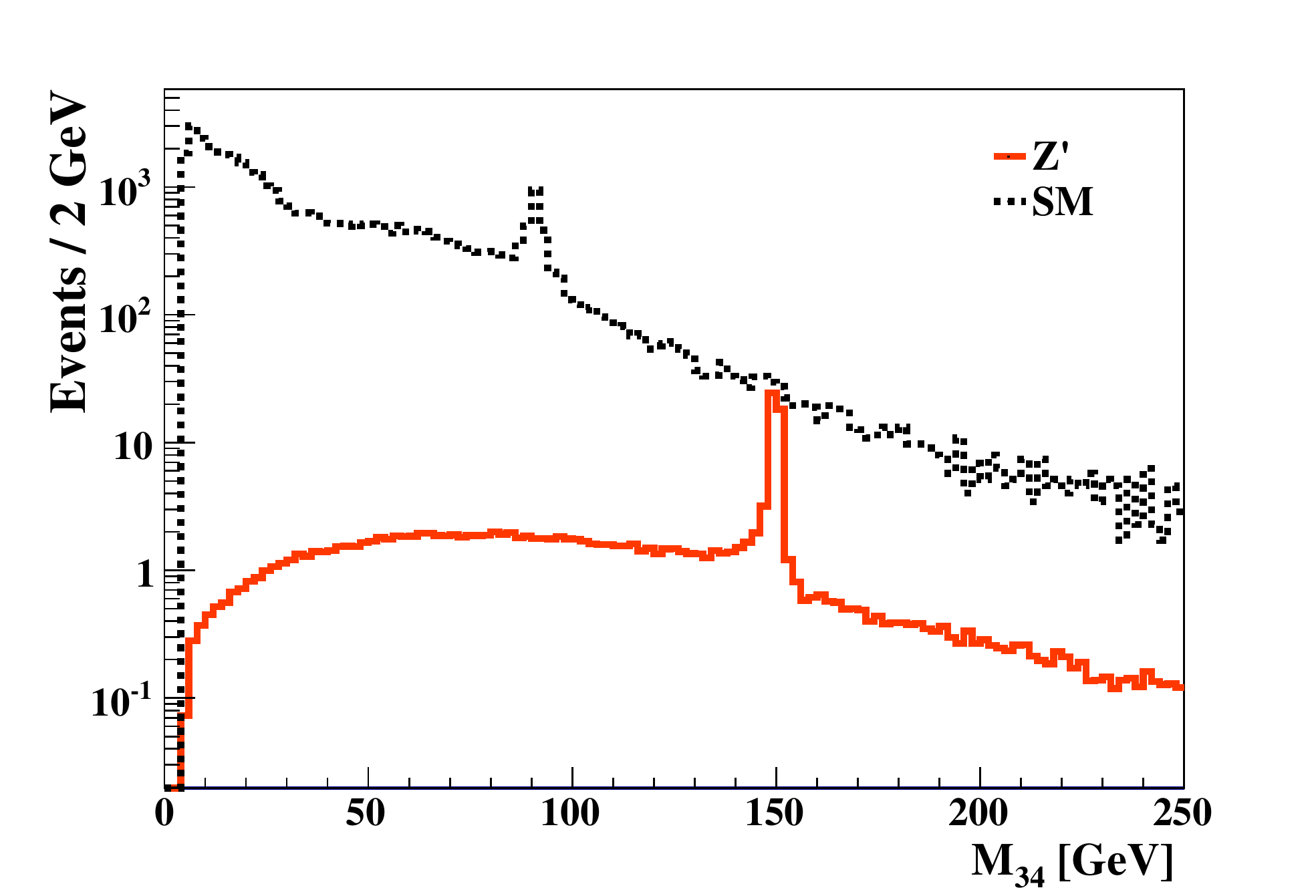}
\end{subfigure}\hfill
\begin{subfigure}
\centering
  \includegraphics[width=\columnwidth]{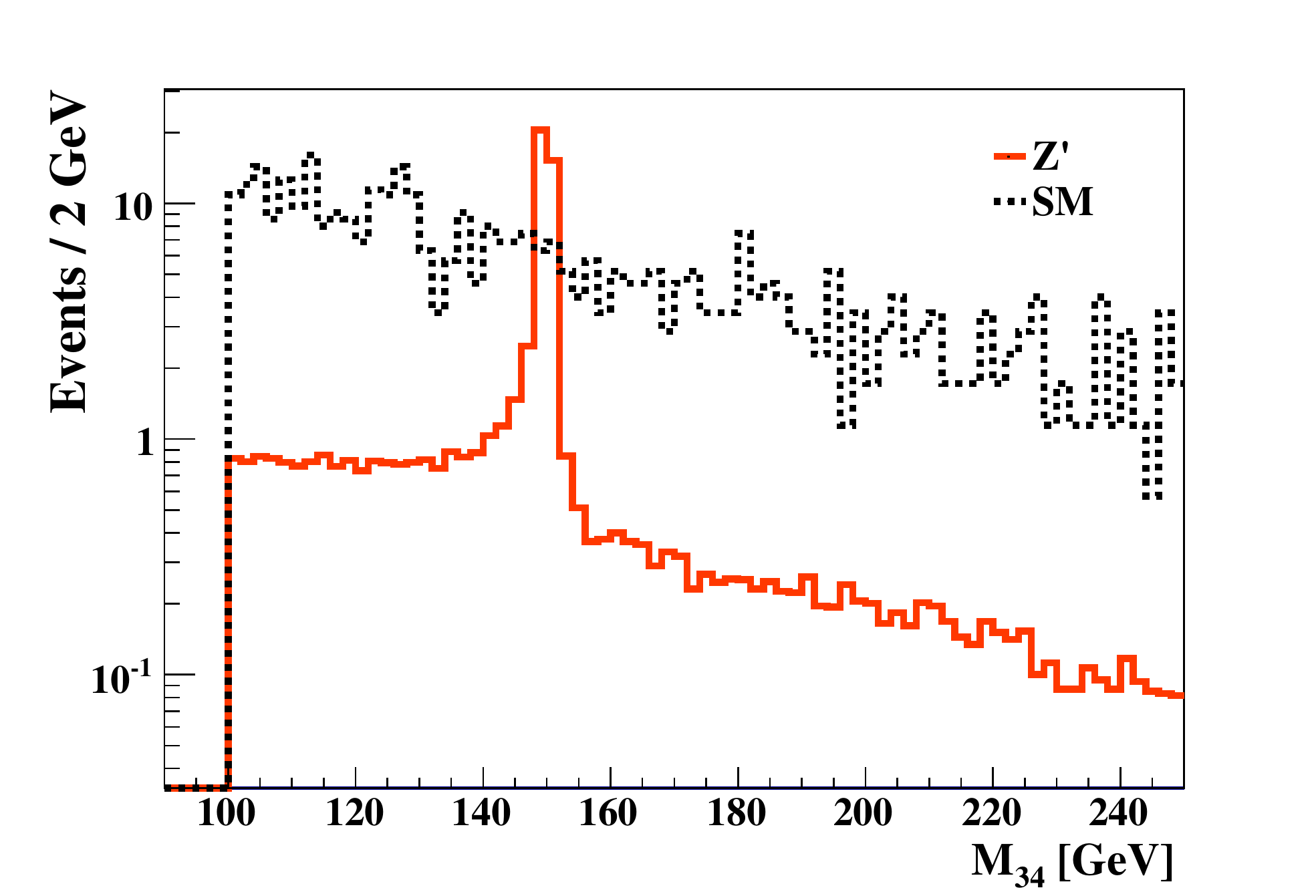}
\end{subfigure}\hfill
\caption{Invariant mass of third and fourth leading in $p_T$ muons before cuts (left) and after cuts (right), for $pp\rightarrow4\mu$ in the SM and $Z'$ model (with $m_{Z'}=150$ GeV, $m_{\chi}=10$ GeV, $g_\mu=g_{\chi}=0.19$), at $\sqrt{s}=14$ TeV and $\mathcal{L}=3000$ $fb^{-1}$.}
\label{fig:tetrmu1}
\end{figure*}

\subsubsection{Four Muon Final State: $m_{Z'}>m_Z$}

We now consider higher $Z'$ masses than those probed by the 4 lepton
search at the $Z$ resonance, i.e., $m_{Z'}>m_Z$.  The $Z'$ production
via the diagram in Fig. \ref{Fig:decay} now proceeds via an off-shell
$Z^*$, with a much lower cross section as seen in
Fig. \ref{Fig:zpcrosssec}.
We perform a similar analysis to that described in the previous
section, with appropriate changes tailored to this higher mass case.
Specifically, we remove the cut on the 4 muon invariant mass, so that
we are no longer restricted to events near the $Z$-resonance, and
place cuts on the di-muon invariant masses to remove the $Z$-peak
(arising from processes in which the $Z'$ in Fig. \ref{Fig:decay} is
replaced by a SM $Z$.)

\begin{figure*}[t]
\centering
  \includegraphics[width=0.82\paperwidth]{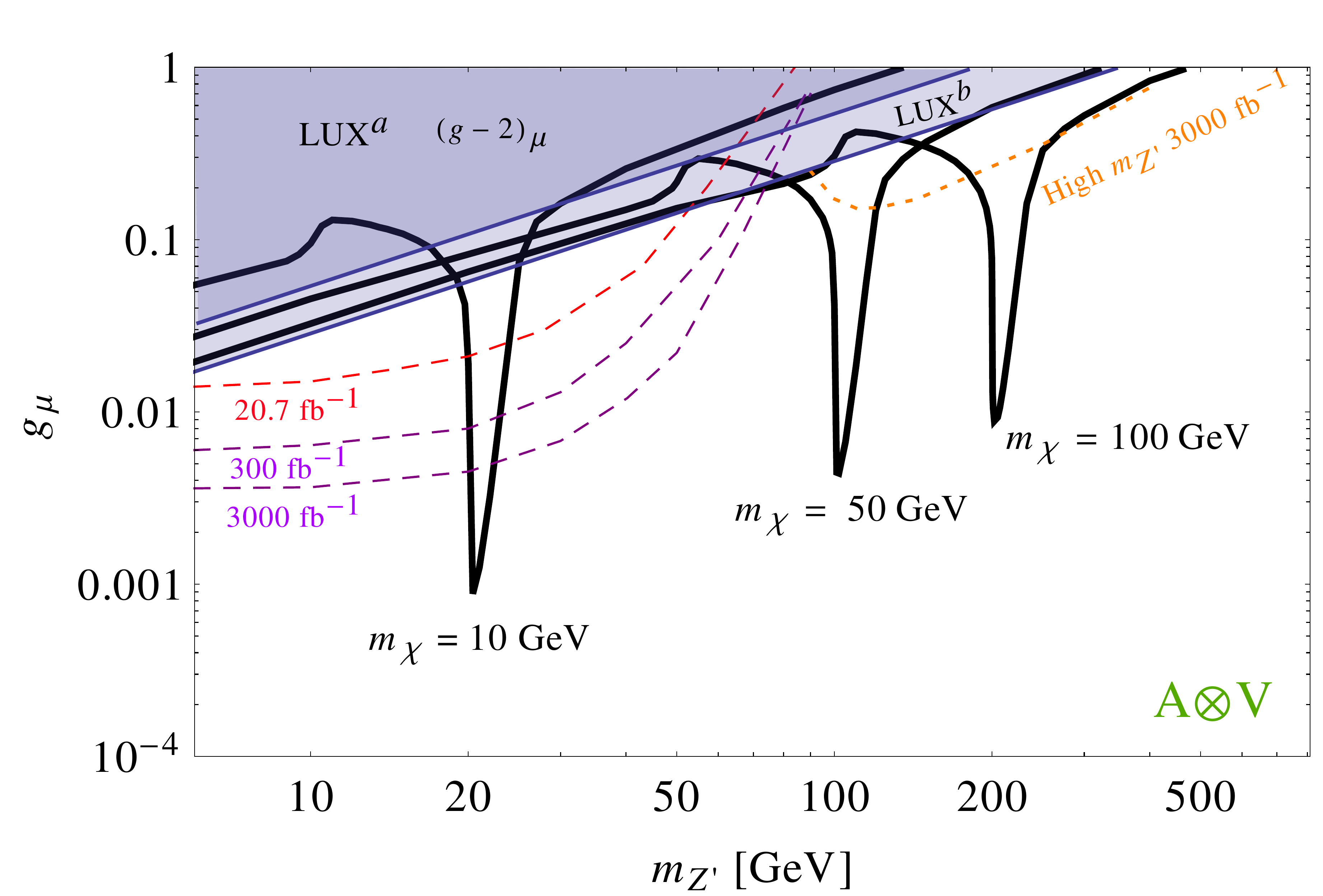}
\caption{Parameter space for $g_\mu$ vs. $m_{Z'}$ with
  $g_\chi=g_\mu$. Shown are excluded regions from $(g-2)_\mu$, as well
  as two excluded regions from LUX direct detection corresponding to
  two different DM mass ranges: LUX$^a$ is for approximately
  $m_\chi=10,1000$ GeV and LUX$^b$ is for the range $m_\chi=20-100$
  GeV. Exclusions from LUX$^a$ and $(g-2)_\mu$ are overlapping in this
  plot.  Relic density curves are shown in black for
  $m_\chi=10,50,100$ GeV.  Dashed lines are ATLAS exclusions and
  reaches: top dashed curve is ruled out by ATLAS data at $\sqrt{s}=8$
  TeV and $\mathcal{L}=20.7$$fb^{-1}$, middle is the ATLAS discovery
  reach at $\sqrt{s}=14$ TeV and $\mathcal{L}=300$$fb^{-1}$, and the
  bottom dashed curve is the ATLAS discovery reach at $\sqrt{s}=14$
  TeV and $\mathcal{L}=3000$$fb^{-1}$.  The dotted line shows the
  ATLAS reach at 3000$fb^{-1}$ for a high mass $Z'$.  The LHC limits
  all assume $m_\chi=10$~GeV.}
\label{fig:mueall22}
\end{figure*}

We implement the following selection cuts:
\begin{itemize}
  \item $p_{T,\mu} > 4$ GeV and $|\eta|<2.7$ for individual muons
  \item Candidate separation of $\Delta R_{\mu\mu}>0.1$ 
  \item $M_{\mu^-,\mu^+} >$ 100 GeV for both the leading pair and next to leading pair in momentum
  \item $p_{T,\mu} >$ 120, 100, 8 GeV for the leading three muons
\end{itemize}
Due to the low cross sections, large luminosities are required to
constrain the high $Z'$ masses.  In Fig. \ref{fig:mueall22} we show the
projected $3\sigma$ exclusion curve at 3000 $fb^{-1}$. Notice that
in the high $m_{Z'}$ searches at high luminosities one is able to probe
$Z'$-masses up to $m_{Z'}\sim 500$ GeV.  We show a
selection of kinematic plots in Fig. \ref{fig:mus3},
\ref{fig:tetrmu1} for an example choice of parameters ($m_{Z'}=150$
GeV, $m_{\chi}=10$ GeV, $g_\mu=g_{\chi}=0.28$) which fall on this
curve.  The effectiveness of the kinematic cuts can be seen by
comparing the LH and RH panels of Fig. \ref{fig:tetrmu1}, which display
the $M_{34}$ distributions before and after cuts, respectively.

\subsection{$Z'$ Decay to Dark Matter}

We now consider the process $pp\rightarrow \ell\ell\chi\chi$ (see
Fig. \ref{Fig:decay}) for which the signal is a pair of opposite sign
leptons plus missing $E_T$.  Unlike the 4 lepton signal, the 2 lepton
+ missing $E_T$ signal is subject to significant backgrounds, which
render the detection prospects very poor.  The dominant SM backgrounds
arise from
\begin{subequations}
 \begin{equation}
pp \rightarrow Z + \textrm{jets} \rightarrow \ell^+\ell^- + \textrm{jets},
 \end{equation}
 \begin{equation}
t\bar{t}\rightarrow b\bar{b}WW\rightarrow b\bar{b}\ell^+\ell^-\nu_{\ell}\bar{\nu}_{\ell}.
 \end{equation}
\end{subequations}
The $Z$ + jets background, with soft jets from the underlying QCD
process, produces an enormous number of dileptons plus missing $E_T$
events, where the missing energy arises from jet misidentification or
energy mismeasurement.  To overcome this background we would need to
select events with $E_T \agt 200$~GeV.  However, a signal in this
range would require a $Z'$ greater than 200 GeV, for which the
production cross section is extremely small.  
Nonetheless, despite being unable to detect the DM production process,
the presence of the $Z'$ coupling to DM still affects the collider
phenomenology through the $Z'$ width and branching fractions, as can
be seen by comparing Fig. \ref{fig:mueall22} and \ref{fig:mueall2}.
Finally, we note that this dilepton plus missing $E_T$ signal could be
a relevant discovery channel in a future lepton collider such as the
ILC, where the $Z$+jets background is not present.

\section{Discussion}
\label{sec:discuss}

Our main results are summarized in Fig.~\ref{Fig:VV} for
$\Gamma_\chi\otimes\Gamma_\ell=V\otimes V$ and
Fig.~\ref{fig:mueall22}, \ref{fig:mueall2}, \ref{fig:mueall} for
$\Gamma_\chi\otimes\Gamma_\ell=A\otimes V$.  Note that because
$V\otimes V$ is highly constrained by the direct detection results, we
have explored the $A\otimes V$ case in much greater detail.

In Fig.~\ref{fig:mueall22}, we summarise the results for a $Z'$ that
couples to the muon flavour.  We see that there is significant overlap
of the excluded regions from $(g-2)_\mu$ and direct detection results,
which each place non-trivial constraints on the model.  In
particular, for values of $m_\chi\lesssim100$ GeV, they rule out those
parameters for which the relic density can be explained, unless the
masses fall in the vicinity of a resonance at $m_{Z'} \sim 2 m_\chi$.
The LHC results place complementary constraints.  For a low mass $Z'$
(those with $m_{Z'} < m_Z$), the collider limits rule out smaller values
of the coupling $g_\ell$ than can be probed by DD or $(g-2)$.  The
projected limits (or discovery sensitivity) for the 14 TeV LHC at
3000~$fb^{-1}$ significantly covers the low $m_{Z'}$ parameter space, even
for parameters for which the DM relic density is controlled by resonant
annihilation.  For a higher mass $Z'$ ($m_{Z'} > m_Z$) the production
cross section at the LHC is suppressed, and hence only the
3000~$fb^{-1}$ results are shown in the figure.  It is clear that, for
sufficiently large $m_\chi$ or $m_{Z'}$, it will be possible to find
parameters which satisfy the relic density requirement and escape all
constraints.

The constraints on a $Z'$ which couples to the $e$ flavour are shown
in LH panel of Fig.~\ref{fig:mueall}.  The DD constraints are similar
for all flavours, as they depend only logarithmically on the lepton
masses.  For electrons, the $(g-2)$ constraint is too weak to be shown
on the plot.  However, the $e$-flavour is subject to LEP-II $Z'$
constraints, which are very strong and eliminate much parameter space
that is open for the $\mu$-flavour.  As a result, the relic density
constraints cannot be met unless one lives very close to a resonance.

Finally, we consider the $\tau$ flavour in the RH panel of
Fig.~\ref{fig:mueall}.  Here the $(g-2)$ constraints are again too
weak to be shown on the plot. As we mentioned above, there are no
current collider analyses for this case, due to the difficulties
associated with tau reconstruction.  Instead, we indicate in
Fig.~\ref{fig:mueall} the $Z'$ production cross section, as a crude
indication of the sensitivity that could be obtained were a dedicated
analysis for the 4-$\tau$ final state to be performed.

In Fig.~\ref{fig:mueall22} and \ref{fig:mueall}, we assumed the $Z'$
couples with equal strength to the DM and leptons, i.e. $g_\chi =
g_\ell$.  If instead we take $g_\chi > g_\ell$, the constraints are
relaxed and thus the allowed region of parameter space enlarged.  This
is illustrated in Fig.~\ref{fig:mueall2}, in which we take
$g_\chi=4g_\mu$ (left panel) and $g_\chi=8g_\mu$ (right panel).  We
see that increased $g_\chi$ lowers the relic density curves.  The LHC
curves are independent of the choice of $g_\chi$ when
$m_{Z'}<2m_\chi$; for $m_{Z'}>2m_\chi$ the constraints on $g_\ell$
become weaker as we enlarge $g_\chi$, due to the increased branching
ratio of the $Z'$ to invisible final states.  In
Figs.~\ref{fig:mueall22} and \ref{fig:mueall2}, this occurs for
$m_{z'} > 20$ GeV, and we note an upturn in the LHC curves at that
point.  The LHC results for high $Z'$ masses $m_{Z'} > m_z$ have not
been shown in Fig.~\ref{fig:mueall2}, as they are weaker than the DD results.

The LHC curves shown in Fig.~\ref{fig:mueall22}, \ref{fig:mueall2} and
\ref{fig:mueall} assume $m_\chi=10$ GeV.  For other DM masses, the
exclusion curves are approximately the same, except for a change to
the point where $m_{Z'} = 2m_\chi$, beyond which the $Z'$ is heavy
enough for the dark branching ratio to be non-zero.\footnote{Even when
  $Br$(DM) is non-zero, if $g_\chi \sim g_\ell$ 
  the cross section and thus the exclusion curves change only
  by a modest factor.}

\vspace{4mm}
\section{Conclusions}
\label{sec:con}

We have considered a leptophilic WIMP scenario in which DM does not
couple to SM quarks at tree-level, and instead couples only to SM leptons.  In
this scenario, the DM has WIMP-scale interaction with leptons,
accounting for the relic density, but suppressed signals in direct
detection experiments and hadron colliders, consistent with
the null results from these searches to date.

We explored such a leptophilic DM in the context of a simple $Z'$
model, in which DM-lepton interactions are mediated by the exchange of
a new vector boson which couples to one of the SM leptons flavours,
$\ell=e$, $\mu$, or $\tau$, with Lorentz structure
$\Gamma_\chi\otimes\Gamma_\ell=V\otimes V$ or $A\otimes V$.  DM-quark
interactions are induced at loop level through kinetic mixing of the
$Z'$ and SM hypercharge, providing nuclear recoil signals in direct
detection experiments. Despite the loop-suppressed nature of this
process, the resulting bounds are strong.  For
$\Gamma_\chi\otimes\Gamma_\ell=V\otimes V$ the DD bounds eliminate all
parameter space for which the correct relic density can be obtained,
except if the DM annihilation cross section has a strong resonant
enhancement.  For $\Gamma_\chi\otimes\Gamma_\ell=A\otimes V$ the
direct detection cross section is velocity suppressed, resulting in
weaker constraints, but even so the DD bounds eliminate significant
parameter space.  For DM coupling to muons, these bounds are
comparable to those from $(g-2)_\mu$.

Production of leptophilic $Z'$ at the LHC occurs via the radiation of
a $Z'$ from Drell-Yan leptons.  We determined exclusion limits for the
$Z'$ mass and coupling strength, using results from a recent ATLAS
analysis of $pp\rightarrow Z\rightarrow 4 e$ or $4\mu$, at
$\sqrt{s}=8$ TeV and $\mathcal{L}=20.7$ $fb^{-1}$.  We also projected
the future exclusion/discovery reach for $\sqrt{s}=14$ TeV and higher
luminosities, for both low and high mass $Z'$ bosons.  
For $\ell=e,\mu$, the combination of the LHC, direct detection, and relic
density constraints excludes most parameter space, except that for
which the DM annihilation at freezeout is resonantly enhanced.  For
$\ell=\mu$, some non-resonant parameter space remains open for $m_\chi \agt 200$
GeV, while for $\ell=e$ this is eliminated by LEP-II bounds.  For
$\ell=\tau$, however, no LHC bounds exist, and much more parameter
space is open.

In conclusion, despite the absence of tree-level interactions with
quarks, this leptophilic dark matter model is strongly constrained by
results from nuclear recoil and hadron collider experiments.

\smallskip

{\it Note added:} A recent paper has calculated limits on a $Z'$ boson
which couples to muon lepton number, based on neutrino trident
production~\cite{Alt:2014, Alt2:2014}.  These limits are a factor of
$\sim 5$ stronger than those from $(g-2)_\mu$.

\vspace{6mm}
\section*{Acknowledgements}
NFB, YC and AM were supported by the Australian Research Council and RKL by the Commonwealth of Australia.

\end{document}